\documentclass[journal]{IEEEtran}
\usepackage{cite}
\usepackage{amsmath,amssymb,amsfonts}
\usepackage{algorithmic}
\usepackage{graphicx}
\usepackage{textcomp}
\usepackage{booktabs}
\usepackage{setspace}
\usepackage{caption}
\usepackage{braket}
\usepackage[hidelinks]{hyperref}

\def\BibTeX{{\rm B\kern-.05em{\sc i\kern-.025em b}\kern-.08em
    T\kern-.1667em\lower.7ex\hbox{E}\kern-.125emX}}
\begin{document}
\bstctlcite{IEEEexample:BSTcontrol}

\title{Scalable Quantum Reservoir Computing over Distributed Quantum Architectures}

\author{Ioannis~Liliopoulos, Georgios~D.~Varsamis, Konstantinos~Rallis, Evangelos~Tsipas, Ioannis~G.~Karafyllidis, Georgios~Ch.~Sirakoulis and~Panagiotis~Dimitrakis
\thanks{This research has been supported by the project “A catalyst for EuropeaN ClOUd Services in the era of data spaces, high-performance and edge computing (NOUS)”, Grant Agreement Number 101135927. Funded by the European Union's HORIZON-CL4-2023-DATA-01 call.}%
\thanks{I. Liliopoulos, G. D. Varsamis, K. Rallis, and E. Tsipas are with the Department of Electrical and Computer Engineering, Democritus University of Thrace, Xanthi, 67100, Greece, and also with the National Centre for Scientific Research "Demokritos", Athens, 15341, Greece. (Corresponding author: Georgios D. Varsamis, e-mail: gevarsam@ee.duth.gr).}%
\thanks{I. G. Karafyllidis and G. Ch. Sirakoulis are with the Department of Electrical and Computer Engineering, Democritus University of Thrace, Xanthi, 67100, Greece.}%
\thanks{P. Dimitrakis is with the National Centre for Scientific Research "Demokritos", Athens, 15341, Greece.}\vspace{-15pt}}

%







\maketitle
\begin{abstract}
Reservoir computing provides an alternative to recurrent neural networks by overcoming the common problems of backpropagation through time and by training only a simple readout layer. The emerging field of quantum computing offers a new computing paradigm that promises to enhance learning through richer feature representations. In this work, we investigate quantum reservoir computing for time-series forecasting. We explore and benchmark four different architectures that combine single or multiple (distributed) reservoirs with single or multiple (distributed) ridge-regression readout layers.  We evaluate these architectures using ideal and hardware-informed noisy simulations, and include both hybrid and fully quantum variants, with classical reservoir counterparts serving as a baseline. The results indicate that quantum-enhanced configurations consistently improve forecasting accuracy {by reducing the mean absolute error (MAE) and the root mean squared error (RMSE) up to 78.8\% and 72.3\%, respectively}, while distributed architectures effectively enable scaling by utilizing multiple quantum resources in a hardware-agnostic manner. These findings support distributed quantum reservoir computing as a promising, modular approach for forecasting on the quantum platforms of the noisy intermediate-scale quantum (NISQ) era.\vspace{-0.5mm}
\end{abstract}


\section{Introduction}
\label{sec1}

Artificial Neural Networks (ANNs) have long been the driving force of the progress in the field of ML. In their general form, they consist of a set of interconnected layers of nodes, the neurons, that are fed with data and undergo a learning process, which allows them to recognize specific patterns. There are several different types and architectures of ANNs, clustered based on their structure and operation, each one with its own special properties and characteristics. Different types of ANNs are capable of solving different types of problems and thus, target different applications. In between the several ANN categories, Recurrent Neural Networks (RNNs), along with their advanced variants such as Long Short-Term Memory (LSTM) and Gated Recurrent Units (GRU), have become foundational models for handling sequential data. By incorporating memory mechanisms, these networks are particularly well-suited for time-series forecasting tasks, where capturing temporal dependencies is critical.

However, the performance of the aforementioned architectures comes at a cost. Their learning process is mainly based on Back propagation Through Time (BPTT) \cite{werbos2002backpropagation}, a process that can become extremely complex with increasing model size and complexity. Additionally, BPTT is susceptible to issues such as vanishing and exploding gradients, which can hinder effective training. The problems of BPTT have driven researchers towards seeking alternative methods to capture complex dynamics.

Derived from RNNs, Reservoir Computing (RC) is a computational framework that consists of two distinct parts. The first is a high-dimensional system, the reservoir, where the connections of existing neurons are kept fixed after initialization, and the second is the readout layer, typically trained by means of a linear or ridge regression rule \cite{cucchi2022hands}. Utilizing the system’s memory introduced by the non-linearity of the randomized connections of the reservoir, responses are generated based on past inputs \cite{schrauwen2007overview, tsakalos2021protein}. This pattern capture and the elimination of the training obstacle, highlights the RC efficiency and simplicity, making them well-suited frameworks for time-series prediction \cite{gauthier2021next,moon2019temporal, wang2019novel}.

Quantum computers are an emerging technology that has demonstrated significant advantages over classical computers in solving certain problems \cite{grover1996fast,shor1999polynomial}. Their intersection with ML, known as Quantum Machine Learning (QML), has been shown to offer advantages by utilizing quantum phenomena such as superposition and entanglement \cite{biamonte2017quantum, mitarai2018quantum, schuld2018supervised}. Several models have been proposed in the literature, including purely quantum or hybrid quantum - classical architectures. Variational Quantum Classifiers (VQCs) \cite{havlivcek2019supervised}, Quantum Neural Networks (QNNs) \cite{beer2020training, farhi2018classification}, Hybrid Classical Quantum NNs (HCQNN) \cite{jin2025ppo,liliopoulos2025hybrid} and Quantum Recurrent Neural Networks (QRNNs) \cite{ceschini2022hybrid, takaki2021learning} are utilizing the variational principle, where through backpropagation, the angles of parameterized rotation gates are optimized to minimize an error function. Quantum Support Vector Machines (QSVMs) \cite{jager2023universal} build upon VQCs and classical SVMs by utilizing quantum kernels. Again, the training process of the aforementioned models can be quite challenging, and, thus there is an active interest in the combination of Quantum computers and RC (QRC) \cite{ahmed2024prediction, ahmed2025optimal, dudas2023quantum, kobayashi2024feedback, suzuki2022natural}.

In this work, we considered the energy - power load forecasting problem, where RC recently showcased great potential \cite{brucke2024benchmarking,guerra2023probabilistic}. We extend established QRC concepts by treating the model as a modular, multi-QPU pipeline, where the corresponding nodes are independent quantum circuits that can be executed in parallel on heterogeneous backends. For the trainable readout layer, we introduce a quantum ridge regression approach \cite{chen2023faster,mohammadisiahroudi2022quantum}, leveraging a trainable quantum kernel \cite{havlivcek2019supervised, schuld2021supervised}. We implement the proposed models using Qiskit \cite{javadi2024quantum, sahin2025qiskit} and NVIDIA CUDA-Q \cite{NVIDIA_CUDA-Q} and compare our results with architecture-matched classical baselines implemented with the ReservoirPy framework \cite{trouvain2020reservoirpy}. Overall, our results show that both distributed QRC (DQRC) and hybrid RC models are able to outperform purely classical RC ones, while remaining scalable via modular partitioning of the reservoir and/or the output layer across multiple QPUs, thereby reducing the per-device qubit requirements and aligning with current NISQ constraints. Importantly, across the evaluated architectures, the quantum-kernel ridge readout layer emerges as the most consistent contributor to forecasting accuracy, whereas component distribution primarily enables effective scaling while also, in some cases, provides additional performance improvements. {Numerical results show that the investigated quantum-enhanced models provide a major boost in forecasting accuracy, achieving peak reductions of up to 78.8\% in mean absolute error (MAE) and 72.3\% in root mean squared error (RMSE) compared to matched classical baselines. }

We were able to achieve this advantage on both ideal simulations and noise model simulations. The simulations were based on actual hardware for the noise models, namely IBM's superconducting ``ibm\_marrakesh" and ``ibm\_brisbane" \cite{IBM_Quantum_2021} as well as IONQ's ion-trap ``Aria-1" \cite{IonQ}, which highlights the versatility and scalability of our approach.

Thus, the rest of the paper is organized as follows: In Section~\ref{sec:architectures} we present and analyze the structure of the reservoir and the output layer as well as the different explored network Architectures. After that, in Section~\ref{sec:results}, we analyze the dataset and the techniques we exploited to evaluate the performance of our models and then compare the efficacy of all models in each Architecture by using three (3) widely used forecasting metrics. Finally, in Section~\ref{sec:conclusions} we summarize the research outcomes of this work and propose future aspects.

\section{Architectures}
\label{sec:architectures}
\begin{figure}[h]
\centering
\includegraphics[width=\columnwidth]{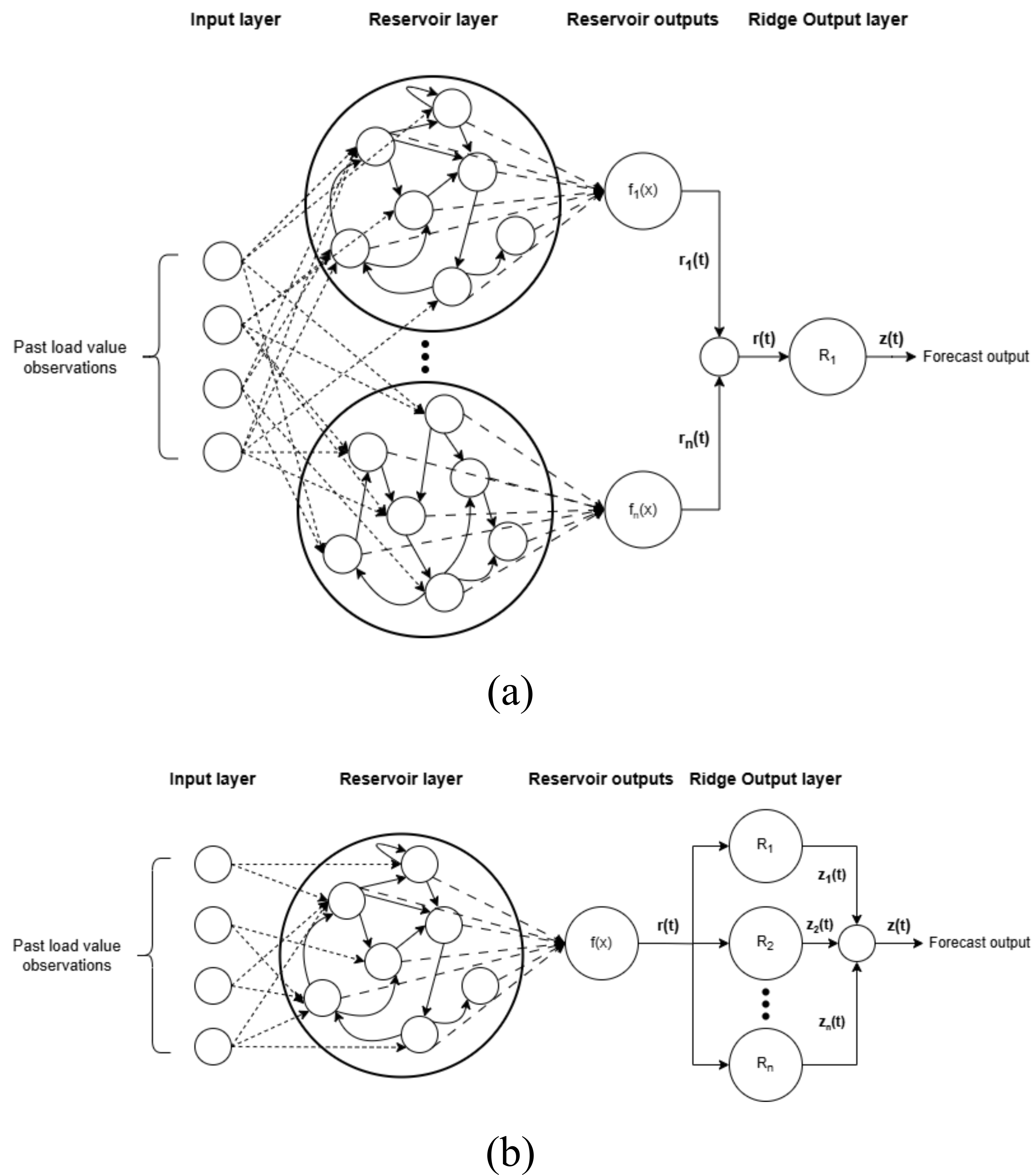}
\caption{(a) Architecture 2 {(MRSR)}, as introduced in Table \ref{table:tab1}, i.e., a RC model consisting of multiple reservoirs and a single output layer. (b) Architecture 3 {(SRMR)}, as introduced in Table \ref{table:tab1}, i.e., an RC model consisting of a single reservoir and multiple ridge instances on the output layer. In both figures $f_i(x)$ or $f(x)$ represents the concatenation of all $i_{th}$ reservoir data into an output vector, without the application of an activation function.}
\label{fig1}
\phantomsection\label{fig1a}
\phantomsection\label{fig1b}
\end{figure}

Within the scope of this work, we developed several scalable quantum reservoir models and validated their performance against a power load forecasting task. Alike most reservoir computing models \cite{bib1, bib2, bib3}, our models comprise a reservoir and an output layer. Unlike previous studies, this work does not emphasize only a single, unified reservoir and output layer RC model structure but is further extended, proposing a modular architecture consisting of multiple reservoir and output layers that can be distributed across different hardware. In particular, in this work we studied four different variations of single and multiple reservoir and output layers, henceforth mentioned as Architectures, as seen in Table \ref{table:tab1}. For illustration purposes, two (2) of the four (4) Architectures, {namely Architecture 2 (the Reservoir-Distributed configuration - MRSR) and 3 (the Readout-Distributed configuration - SRMR) are depicted in Figs. (\ref{fig1a}a) and (\ref{fig1b}b) respectively.} Regarding Architectures 1 and 4, their structural diagrams are omitted here and are provided in Appendix \ref{appA}, for readability purposes.

\begin{table}[h]
\caption{Reservoir Architecture details}\label{table:tab1}%
\resizebox{\columnwidth}{!}{%
\begin{tabular}{c c}
\toprule
 No.  & Details\\
\midrule
1 & Single Reservoir with Single Ridge Regression Kernel (SRSR)\\
2 & Multiple Reservoirs with Single Ridge Regression Kernel (MRSR)\\
3 & Single Reservoir with Multiple Ridge Regression Kernels (SRMR)\\
4 & Multiple Reservoirs with Multiple Ridge Regression Kernels (MRMR)\\
\bottomrule
\end{tabular}
}
\end{table}

Additionally, as illustrated in Figs.~\ref{fig1a}a and \ref{fig1b}b, each reservoir neuron can receive inputs from both input data and other reservoir neurons. The number of inputs that each reservoir neuron can receive is fixed and equal to 4. For each $i_{th}$ neuron in the reservoir, up to two of its inputs ($k_i \in \{0,1,2\}$) are signals selected from other reservoir neuron outputs, while the remaining $4-k_i$ inputs are taken directly from the input data. The value of each neuron $k_i$ is randomly chosen for each neuron during initialization. This approach implies that the data have to be repeatedly propagated through the neurons of the neural network so that all neurons are properly activated. In the present study, we selected to propagate our data three times within the reservoir layer to ensure that all neurons are activated properly.

\begin{figure*}[t!]
\centering
\includegraphics[width=\textwidth]{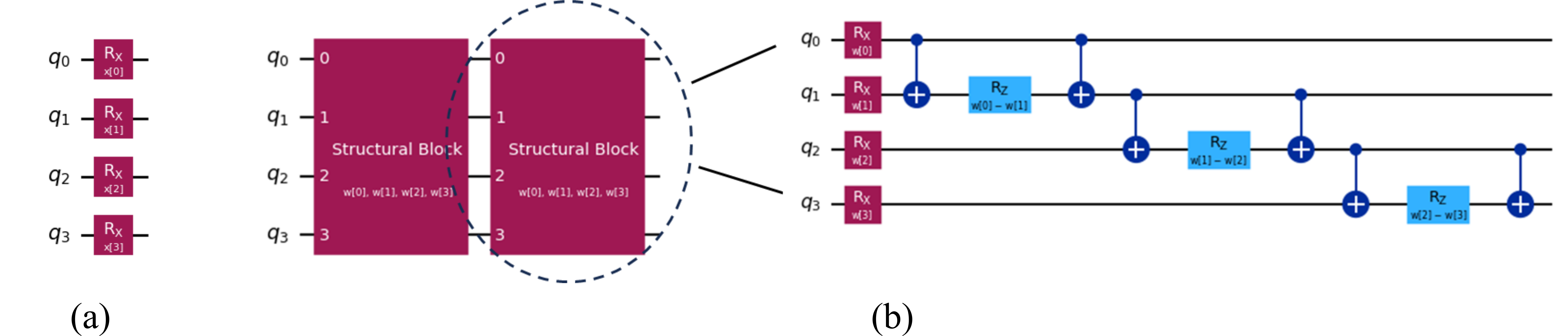}
\caption{The Feature Map $\Phi$ used for each quantum neuron. We applied a $R_x$ gate to each qubit, and the rotation angle $\theta$ is equal to the input data for the corresponding neurons. (b) The Ansatz for each quantum neuron, comprising two structural blocks. Each block is a four-qubit quantum circuit consisting of $R_x$ and $R_z$ gates with trainable $w_i$ parameters and controlled-NOT gates.}
\label{fig2}
\phantomsection\label{fig2a}
\phantomsection\label{fig2b}
\end{figure*}

\subsection{Quantum reservoir layer}
\label{subsec:reservoir_layer}
More specifically, the reservoir layer we designed consists of one or more N-neuron reservoirs. Each neuron within those reservoirs is a 4-qubit quantum circuit, and thus we will refer to it as ``quantum neuron". Each quantum neuron consists of two parts, the Feature Map circuit and the Ansatz. For the Feature Map circuit, we exploited a simple quantum circuit consisting of rotational gates around the X axis, denoted as $R_X(x_i)$, as demonstrated in Fig.~\ref{fig2a}a. These gates follow the expression $R_X(x_i) = exp(-i\frac{x_i}{2}\sigma^x)$, where $\sigma^x$ is the Pauli X operator, and are instrumental in the data encoding process since they are responsible for the conversion of the classical variables into rotation angles, applied to each qubit, a technique called ``angle encoding" \cite{bib4}.

Inspired by the work of Suzuki et al. \cite{bib5}, we exploited the circuit demonstrated in Fig.~\ref{fig2b}b as the Ansatz of each quantum neuron. This circuit comprises rotational gates around the $X$ and $Z$ axes and Controlled-NOT (CNOT) gates. The mathematical expression of the $R_Z(w_i)$ gate is $R_Z(w_i) = exp(-i\frac{w_i}{2}\sigma^z)$, where $\sigma^z$ is the Pauli Z operator, and thus the ansatz structure may be expressed as follows:



\begin{align*} 
U_{ansatz} &= \prod_{k=1}^{B} \Bigg[ \prod_{j=0}^{n-1} exp(-iw_j\sigma^x)  \\
&\hspace{-1em}\times
\prod_{i=1}^{n} CX_{i,i-1}exp(-i(w_i-w_{i-1})\sigma^z)CX_{i,i-1} \Bigg]  \tag {1}
\end{align*}

where term $B$ refers to the number of structural blocks in the ansatz and $CX_{i,i-1}$ denotes the application of the CNOT gate between qubits $i$ and $i-1$, considering the $i-1$ qubit as the control and the $i$ qubit as the target qubit. Both the feature mapping circuit and the ansatz are hardware-efficient quantum circuits since they consist of CNOT and Pauli rotational gates, i.e., gates that are easily realized in the existing quantum hardware. 
In this work, we considered the estimated expectation value $\braket{O}_{\psi_i}$ of each quantum neuron $i$ as its output, by using the Pauli Z gate as the target observable $O$:

\begin{align*} 
\braket{O}_{\psi_i} &= \bra{\psi_i(x,w)}O\ket{\psi_i(x,w)} \\
&= \bra{\psi_i(x,w)}Z\ket{\psi_i(x,w)} \tag {2}
\end{align*}

\subsection{Quantum output layer}
In most reservoir computing models the output layer consists of a fully-connected Neural Network. Training of this network's parameters is obtained using supervised learning, in terms of either a linear or a ridge regression rule \cite{cucchi2022hands}. The main purpose of this layer is to map the reservoir outputs $r$ to the target signal $z$ via an appropriate transformation $W_{out}$:

\begin{align*} 
z(t) = W_{out}r(t) \tag {3} \label{eq3}
\end{align*}
In our work, since we exploit more than a single reservoir in some of our model Architectures, the term $r(t)$ denotes a single vector comprising all reservoir outputs, vertically concatenated:
\begin{align*} 
r(t) =  \begin{bmatrix}
r_1(t) \\
r_2(t) \\
\vdots \\
r_n(t)
\end{bmatrix}
\tag {4}
\end{align*}
Before the training process, both the desired outputs $y(t)$ and the concatenated reservoir outputs $r(t)$ are aggregated into an output vector $Y$ and a single vector $R$, accordingly. In the classical machine learning models, that we used as baseline for our quantum reservoir models, the matrix $W_{out}$ is computed by using the ridge regression method, as follows:

\begin{figure*}[t!]
\centering
\includegraphics[width=\textwidth]{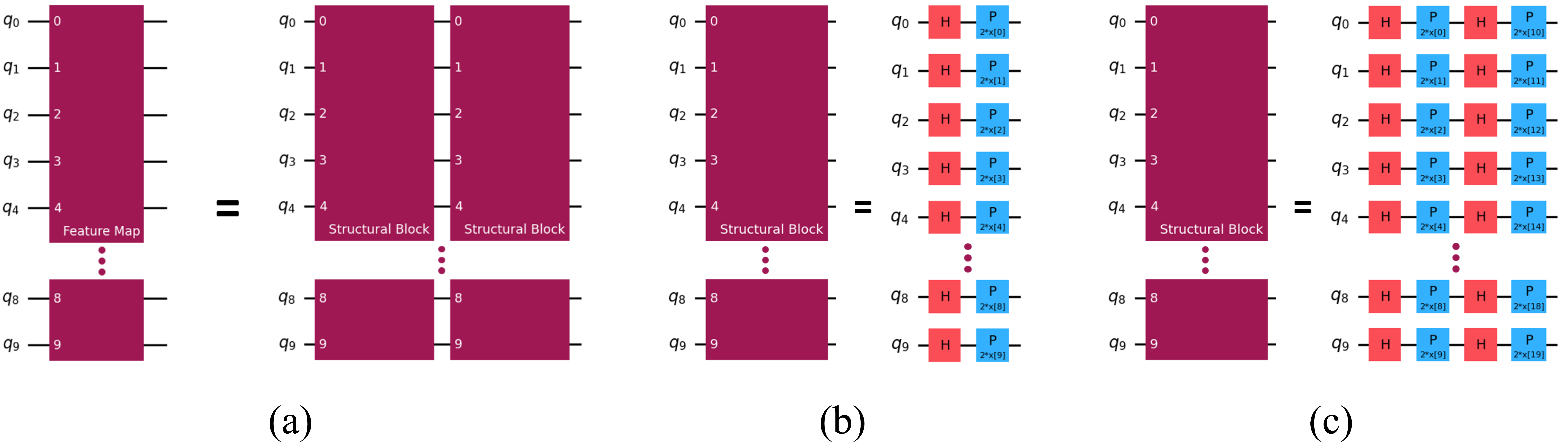}
\caption{(a) The Feature Map used for the quantum kernel output, which comprises two structural blocks. (b) Each structural block is a $n$-qubit quantum circuit, which consists of Hadamard and Phase gates, applied alternatively on each qubit. The input data are encoded via Phase gates. (c) If the input parameters outnumber the available qubits in the feature map, the quantum circuit is adjusted accordingly by adding supplementary layers to reflect the number of input parameters. In this case, for example, $n = 10$ while the number of input parameters is equal to 20; therefore, two (2) layers of Hadamard-Phase gates are needed to properly encode them.}
\label{fig3}
\end{figure*}

\begin{align*} 
\hat{W}_{out} = YR^T(RR^T + \lambda I)^{-1} \tag {5}
\end{align*} 
whereas the term $\lambda$, with $\lambda > 0$, refers to the ridge regularization parameter \cite{khalaf2005}, used to prevent overfitting during the network's training.

Unlike previous studies, in this work, we propose the use of a quantum kernel-based ridge regression, instead of a classical one, for the output layer of all our models. Thus, our models consist of both quantum reservoirs and quantum output layers, leading to fully quantum modular distributed architectures. More specifically, the kernel function $K$, which maps the reservoir layer outputs to the model's outputs, similarly to the $W_{out}$ parameter in Eq. \ref{eq3}, is defined as the overlap of two quantum states produced by a parametrized quantum circuit $\Phi$, known as the feature map circuit:

\begin{align*} 
K(R,Y) = |\braket{\Phi{(R)} | \Phi{(Y)} }|^{2} \tag {6} \label{eq6}
\end{align*} 
Herein, we exploited a custom $n$-qubit feature mapping circuit comprising Hadamard and Phase gates, as depicted in Fig. \ref{fig3}. In Architectures 1 and 2, the number of qubits, $n$, was set to a fixed value of 10. On the other hand, in Architectures 3 and 4, $n \in \{5,10\}$, with respect to the total number of reservoir neurons, in each model. More specifically, when the total number of reservoir neurons is 10, 15 and 25, respectively, $n$ was set to 5, resulting in a distributed output layer consisting of 2, 3 and 5 output ridge instances, respectively. In all other cases $n$ was set to 10. Similarly to the quantum neuron ansatz, the output layer feature map consists of two structural blocks as well. Therefore, in Eq. \ref{eq6}, $\Phi$ is defined as:


    

\begin{align*} 
\Phi = \prod_{k=1}^{B} \prod_{j=1}^{L}\left[ H^{\otimes n}\prod_{i=1}^{n} exp(ix_i) exp(-i(2*x_i)\sigma^z) \right]  \tag {7}
\end{align*} 
since, from \cite{mckay2017efficient}, it is known that the phase gate can be expressed as 
\begin{align*} 
P(\theta) = e^{i\theta/2}RZ(\theta) =  e^{i\theta/2} e^{-i\theta\sigma^z} \tag {8}
\end{align*} 
Additionally, terms $L$ and $B$ correspond to the number of layers in each structural block and to the number of structural blocks in the feature map circuit, respectively.

\label{subsec:arch1}

\section{Results}
\label{sec:results}
\subsection{Dataset}

To test and validate the performance and forecasting capabilities of our models, we created a power load forecasting dataset comprising hourly load observations in the Greek region; these data are publicly available and accessible through the IPTO data portal \cite{IPTO}. More specifically, this dataset consists of 35,064 hourly load values (approximately 49 months). We divided this dataset into smaller ones that were used to train, evaluate, and test our models; The training dataset comprises 24,539 discrete values, whereas the evaluation and testing ones consist of 5,258 and 5,267 values, respectively.

Like every other reservoir computing model, a set of inputs must be introduced to the reservoir to be mapped in a higher-dimensional space. Then, these high-dimensional reservoir outputs are provided to the readout layer for pattern analysis to be performed. In particular, in time-series forecasting, a set of past observations is introduced to the reservoir model to predict a future value. In this research work, we exploited a set of 4 past-hour observations to predict the next-hour load value, measured in MW, as demonstrated in Table \ref{table:tab2}. This technique is commonly referred to in the literature as the "sliding window approach" and is used to express the forecasting problem in terms of a supervised machine learning problem \cite{gasparin2022deep}. 

\begin{table}[h]
\centering
\caption{Four-hour intervals of past observations and the future load value, as exploited in this work, before normalization.}\label{table:tab2}%
\resizebox{\columnwidth}{!}{%
\begin{tabular}{ c c c c c}
\toprule
 \multicolumn{4}{c}{Past Load Observations} & Future Load Value\\
 hour-3  & hour-2 & hour-1 & current hour & \\
\midrule
4,182 & 3,899 & 3,932 & 3,945 & 3,795 \\
3,899 & 3,932 & 3,945 & 3,795 &	3,911 \\
\vdots &\vdots &\vdots &\vdots &\vdots\\
5,702 & 5,703 & 5,320 & 4,814 &	4,414 \\
\bottomrule
\end{tabular}
}
\end{table}

Furthermore, to ensure that our models are able to capture all patterns and trends within our load data \cite{bib6}, we normalized all load values using the Min-Max normalization technique before introducing them to our models:

\begin{align*} 
x_{norm} = \frac{x - x_{min}}{x_{max} - x_{min}} \tag {9}
\end{align*} 

\noindent whereas $x_{min}$ and $x_{max}$ are the minimum and maximum load values, and $x$ is the current load value in each of our datasets accordingly. Therefore, all load values are in the range $[0,1]$. This technique also allowed us to deal with exploding gradient values, which might appear throughout our models' training phase, resulting in improving their overall performance.

\subsection{Performance}
\label{subesec:performance}
We evaluated the performance of our models by exploiting two widely used first- and second-order metrics in time series forecasting, namely the mean absolute error (MAE) and the root mean squared error (RMSE), due to their simplicity of computation and their overall good ability to assess our models' forecasting capabilities:

\begin{align*} 
MAE & = \frac {1}{m}\sum _{i=1}^{m}\mid y_{i} - \hat y_{i}\mid \tag {10}\\
RMSE & = \sqrt {\frac {1}{m}\sum _{i=1}^{m}(y_{i} - \hat y_{i})^{2}} \tag {11}\\
\end{align*} 

Here, $\hat{y}_i$ and $y_i$ represent the predicted and the true values accordingly, and $m$ is the sample size. In addition to these metrics, we also utilized the $R^2$ score, as a secondary metric, to evaluate the efficacy of our models:

\begin{align*} 
R^2 = 1 - \frac{\sum_{i=1}^{n} (y_i - \hat{y}_i)^2}
{\sum_{i=1}^{n} (y_i - \bar{y})^2} \tag {12}
\end{align*} 
where $\bar{y}$ stands for the mean of the true values. The combination of these three (3) metrics allowed us to conduct a comprehensive performance evaluation of our models.

For each evaluated architecture and each different configuration of sub-reservoirs and 
ridge output instances, we consider as a baseline the corresponding fully classical model, where both the reservoir layer (single or distributed) and the ridge output layer (single or distributed) are implemented classically, without the involvement of any quantum components. All hybrid and fully quantum variants are reported alongside this architecture-matched classical-classical baseline to enable a fair comparison. Also, a more visual comparison with the corresponding baselines can be seen in the figures with the results, as presented for every different Architecture (Fig.~\ref{fig4}, Fig.~\ref{fig5}, Fig.~\ref{fig6} and Fig.~\ref{fig7}).

For all our experiments, we executed our quantum models both on ideal and on noisy simulators, able to simulate real quantum device capabilities and existing noise, by using real-time quantum processing unit (QPU) calibration data provided through the Qiskit and NVIDIA CUDA-Q frameworks. Using this approach, we were able to assess the performance of our models with and without the presence of up-to-date quantum system noise. More specifically, we exploited, as mentioned in Section~\ref{sec1}, IBM's superconducting ``ibm\_marrakesh" and ``ibm\_brisbane" as well as IONQ's ion-trap ``Aria-1" QPU system calibration data for our noise models. For reproducibility purposes, we included a table regarding the calibration data values at the time of testing for all these three QPUs in Appendix \ref{appB}.
As mentioned above, three (3) of the four (4) Architectures considered comprise multiple reservoirs or multiple ridge instances. In our tests, for these cases, we considered reservoir layers consisting of 2, 3 and 5 reservoirs. The same applies for the output layer also, where the number of ridge instances was set to 2, 3, and 5, respectively.
Therefore, each noise model is assigned one or two quantum reservoirs, as presented in Table \ref{table:tab3}. The same reasoning also extends to the quantum output layer. For instance, consider a fully quantum model consisting of three reservoirs in the reservoir layer and a single ridge instance in the output layer to be executed on a noisy simulator. According to Table \ref{table:tab3}, the three (3) reservoirs will be evenly distributed to the three available QPU systems to be executed simultaneously, and the ridge instance quantum circuit will be executed on the `IBM Brisbane' quantum device.

\begin{table}[h]
\caption{Quantum Reservoir/Ridge instance distribution among available QPU systems}\label{table:tab3}
\centering
\resizebox{\columnwidth}{!}{%
\begin{tabular}{c c c c}
\toprule
\shortstack{Number of Reservoirs/\\Ridge Instances} & \shortstack{IBM\\Marrakesh} & \shortstack{IBM\\Brisbane} & \shortstack{IONQ\\Aria-1}\\
\midrule
1 & - & 1 & - \\
2 & 1 & 1 & - \\
3 & 1 & 1 & 1 \\
5 & 2 & 2 & 1 \\
\bottomrule
\end{tabular}
}
\end{table}

In what follows, we will present and analyze the results from our tests for each Architecture in a separate subsection.

\subsubsection{Architecture 1: Single Reservoir Single Readout (SRSR)}
\label{sec:subsec321}
Models that fall into this category consist of a single reservoir and output layer, as shown in Fig.~\ref{figA1a}a in Appendix \ref{appA}. For our tests, we considered reservoirs that consist of {10,15,20,25,30,40,50, and 60} neurons accordingly. For this Architecture, the noisy results were obtained by executing the corresponding quantum parts of our models in `IBM Brisbane' as seen in Table \ref{table:tab3}. In Table \ref{table:tabArc1tot} only the best overall performing models are demonstrated with respect to the number of reservoir neurons. When a quantum model exhibits the strongest performance, for any number of neurons, the type of execution is also mentioned in the ``reservoir type" and ``ridge type" columns, noted as (Sim.) for ideal simulations and as (Noisy) for simulations that incorporate noise models. This convention is adhered to throughout the rest of this work. For a detailed performance overview of all models of Architecture 1, refer to Appendix \ref{appC}, in Tables \ref{table:tabC1} - \ref{table:tabC3}:

\begin{table}[h]
\caption{Architecture's 1 {(SRSR)} best-performing model in accordance with the number of reservoir neurons.}
\label{table:tabArc1tot}
\addtolength{\tabcolsep}{-0.26em}
\resizebox{\columnwidth}{!}{%
\begin{tabular}{c|c|c|ccc}
\hline
\textbf{Number} & \textbf{Reservoir Type} & \textbf{Ridge Type}
 & MAE & RMSE & $R^2$ \\
 \textbf{of Neurons} & & & & &\\
\hline
10  & Quantum (Sim.) & Quantum (Sim.) & 0.0141 & 0.0198 & 0.9850\\
20 &  Quantum (Sim.) & Quantum (Sim.) & 0.0141 & 0.0201 & 0.9845\\
40 & Classical & Quantum (Sim.)  & 0.0141 & 0.0204 & 0.9841\\
15  & Quantum (Sim.) & Quantum (Sim.) & 0.0141 & 0.0204 & 0.9841\\
30 &  Classical & Quantum (Sim.) & 0.0142 & 0.0195 & 0.9854 \\
60 &  \textbf{Classical} & \textbf{Quantum (Sim.)}  & \textbf{0.0116} & \textbf{0.0165} & \textbf{0.9896} \\
25 &  Quantum (Sim.) & Quantum (Sim.)  & 0.0139 & 0.0197 & 0.9851 \\
50 &  Quantum (Sim.) & Quantum (Sim.)  &  0.0141 & 0.0202 & 0.9840  \\
\hline
\end{tabular}
}
\end{table}



\begin{figure*}[h!]
\centering
\includegraphics[width=0.7\textwidth]{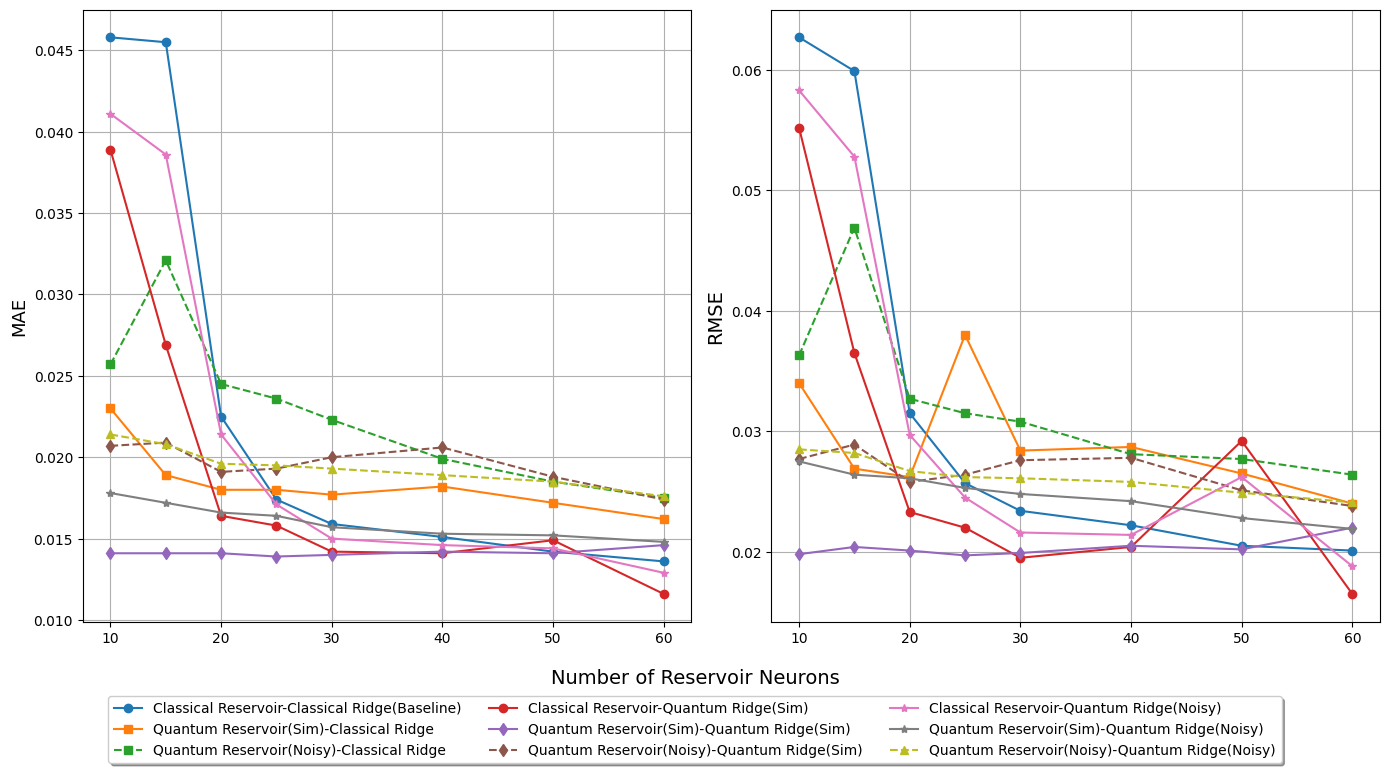}
\caption{Mean Absolute Error and Root Mean Square Error distribution for all Architecture 1 (SRSR) models considered.}
\label{fig4}
\end{figure*}

From Table \ref{table:tabArc1tot} it is evident that all models exploiting the quantum ridge as an output layer yield the best forecasting results overall. {This SRSR configuration yields an average MAE reduction of 28.5\% across all cases, with a peak reduction value of 69.2\% for smaller reservoirs, where classical models typically struggle. A similar behavior is also viewed regarding the RMSE, with an average reduction of 29.8\%, peaking at 68.5\%, indicating an increase in the reliability of the network.} In addition to Table \ref{table:tabArc1tot}, we included Fig.~\ref{fig4} where the MAE and RMSE variance are presented with respect to the number of neurons in the reservoir. In this figure, one may observe that even with a small number of neurons in the reservoir, the quantum models are able to yield quite good performance compared to the hybrid or the classical ones.

\subsubsection{Architecture 2: Multi-Reservoir Single-Readout (MRSR)}
\label{sec:subsec322}
{To address the hardware constraints and qubit limits that are present in the aforementioned monolithic systems, a new Architecture is introduced, which uses a distributed reservoir layer where feature extraction is partitioned across multiple independent quantum nodes before being aggregated into a single readout layer.} Thus, models that fall into this category consist of multiple reservoirs and a single output layer, as presented in Fig. (\ref{fig1a}a). More specifically, the number of neurons per reservoir was selected so as to ensure that the total number of neurons corresponds to the ones of the models that fall into Architecture 1 (SRSR). For example, to match the model case with the 20 neurons in the reservoir, a model belonging in this Architecture with 2 reservoirs shall have 10 neurons in each reservoir. For this Architecture, we used all three available noisy QPU-based simulators, as seen in Table \ref{table:tab3}, to obtain the noisy results. Like in the Architecture 1 subsection, Table \ref{table:tabArc2tot} presents only the best overall-performing models with respect to the number of reservoirs and reservoir neurons. In this table, the abbreviation ``Res-Num $\times$ Res-Neur" indicates the number of reservoirs in the reservoir layer and the number of neurons comprising each reservoir, respectively. For a detailed performance overview of all models that belong to Architecture 2 (MRSR), refer to Appendix \ref{appC}, in Tables \ref{table:tabC4} - \ref{table:tabC6}.

\begin{table}[h]
\caption{Architecture 2 {(MRSR)} best-performing model in accordance with the number of reservoirs in the reservoir layer and the number of neurons comprising each reservoir.}
\label{table:tabArc2tot}
\addtolength{\tabcolsep}{-0.26em}
\resizebox{\columnwidth}{!}{%
\begin{tabular}{c|c|c|ccc}
\hline
\textbf{Res-Num x} & \textbf{Reservoir Type} & \textbf{Ridge Type}
 & MAE & RMSE & $R^2$ \\
\textbf{Res-Neur} & & & & &\\
\hline
2x5  & Quantum (Sim.) & Quantum (Sim.) & 0.0139 & 0.0196 & 0.9853\\
2x10 &  Quantum (Sim.) & Quantum (Sim.) & 0.0140 & 0.0198 & 0.9850\\
2x20 & Quantum (Sim.) & Quantum (Sim.)  & 0.0140 & 0.0202 & 0.9845\\
3x5 & Quantum (Sim.) & Quantum (Sim.) & 0.0142 & 0.0199 & 0.9848\\
3x10 & Quantum (Sim.) & Quantum (Sim.) & 0.0141 & 0.0204 & 0.9841 \\
3x20 &  \textbf{Classical} & \textbf{Quantum (Sim.)}  & \textbf{0.0113} & \textbf{0.0158} & \textbf{0.9904} \\
5x5 &  Quantum (Sim.) & Quantum (Sim.)  & 0.0142 & 0.0202 & 0.9844 \\
5x10 &  Quantum (Sim.) & Quantum (Sim.)  &  0.0143 & 0.0211 & 0.9830  \\
\hline
\end{tabular}
}
\end{table}


\begin{figure*}[!h]
\centering
\includegraphics[width=0.7\textwidth]{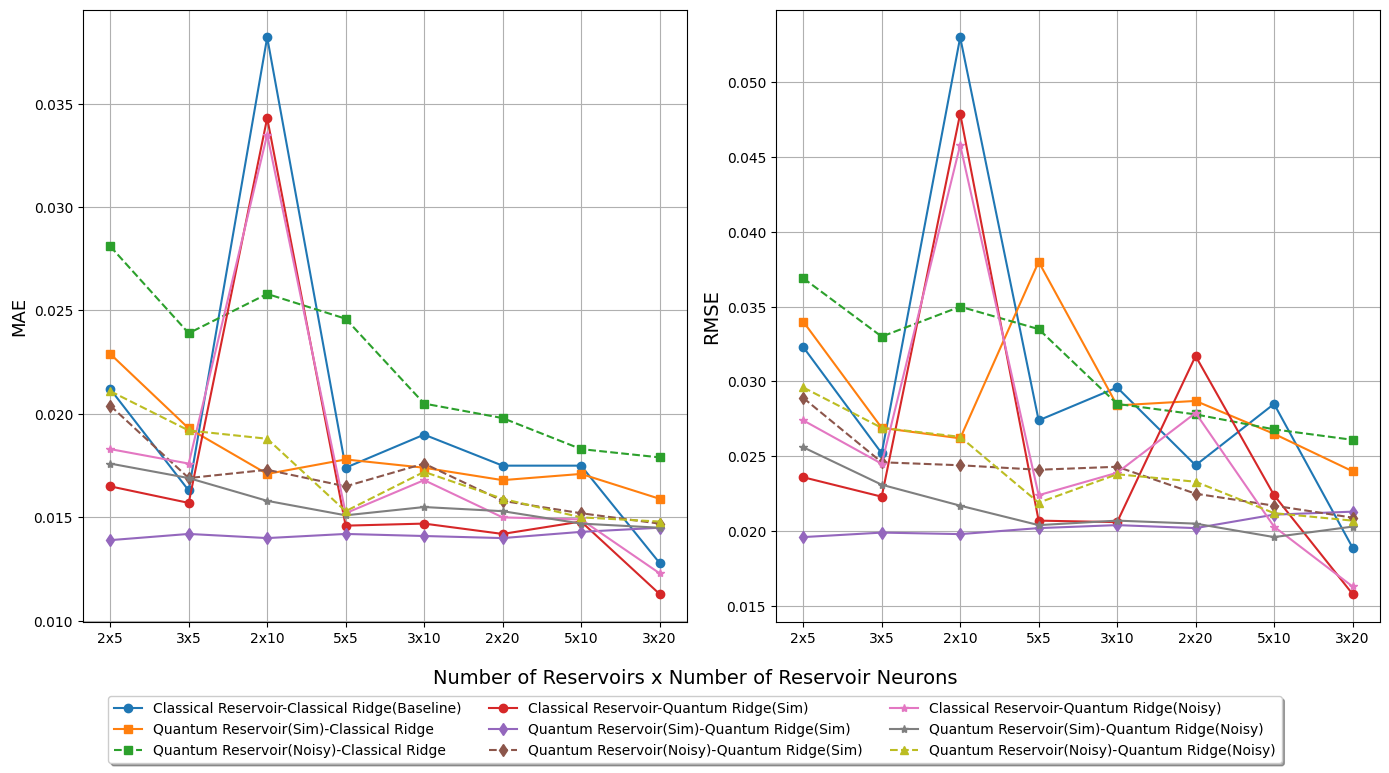}
\caption{Mean Absolute Error and Root Mean Square Error distribution for all Architecture 2 (MRSR) models considered.\vspace{-10pt}}
\label{fig5}
\end{figure*}

The results presented in Table \ref{table:tabArc2tot} indicate that the adaptation of a quantum instead of a classical output layer leads to a greater performance of all models regardless of the type of reservoir. {Beyond supporting hardware scalability, the MRSR Architecture demonstrates significant forecasting gains, achieving an average improvement of 25.6\% in MAE and 30\% in RMSE over the classical distributed reservoir baseline.} Additionally, one may observe that the exploitation of a distributed quantum reservoir along with a quantum ridge output layer yields notable results even with a small number of neurons in each reservoir, as presented in Fig.~(\ref{fig5}), {providing a peak improvement in performance compared to the regarding baseline for the case of 2 quantum reservoirs with 10 neurons each and a ridge output (second line of Table~\ref{table:tabC5}, with 63.3\% lower MAE and 62.6\% lower RMSE}. However, the best outcomes, in this Architecture, were provided from a hybrid model comprising 3 reservoirs with 20 neurons each alongside a quantum output layer. In addition to that, this model's performance is the best among all models comprising a single output layer, i.e., models belonging in Architectures 1 and 2.

\subsubsection{Architecture 3: Single-Reservoir Multi-Readout (SRMR)}
\label{sec:subsec323}
In the previous two subsections, we analyzed models which consist of a single output layer. {Shifting the focus from reservoir partitioning to output scalability, another Architecture is introduced that investigates the impact of distributing the readout process itself by feeding the outputs of a unified reservoir into a parallelized, multi-instance quantum ridge-regression layer.} 
More specifically, in this Architecture, all models consist of a single reservoir and multiple ridge output instances, either classical or quantum, as shown in Fig. (\ref{fig1b}b). In both the reservoir and output layers, for equal comparison purposes, the number of neurons and qubits are the same as in Architectures 1 and 2 {(SRSR and MRSR, respectively)}. Similarly to the previous two (2) Architectures, Table \ref{table:tabArc3tot} presents only the best overall performing models with respect to both the number of reservoir neurons and the number of ridge instances. We also include Tables \ref{table:tabC7} - \ref{table:tabC9} in Appendix \ref{appC}, where all the forecasting results of the models tested are presented in detail.



\begin{figure*}[!h]
\centering
\includegraphics[width=0.7\textwidth]{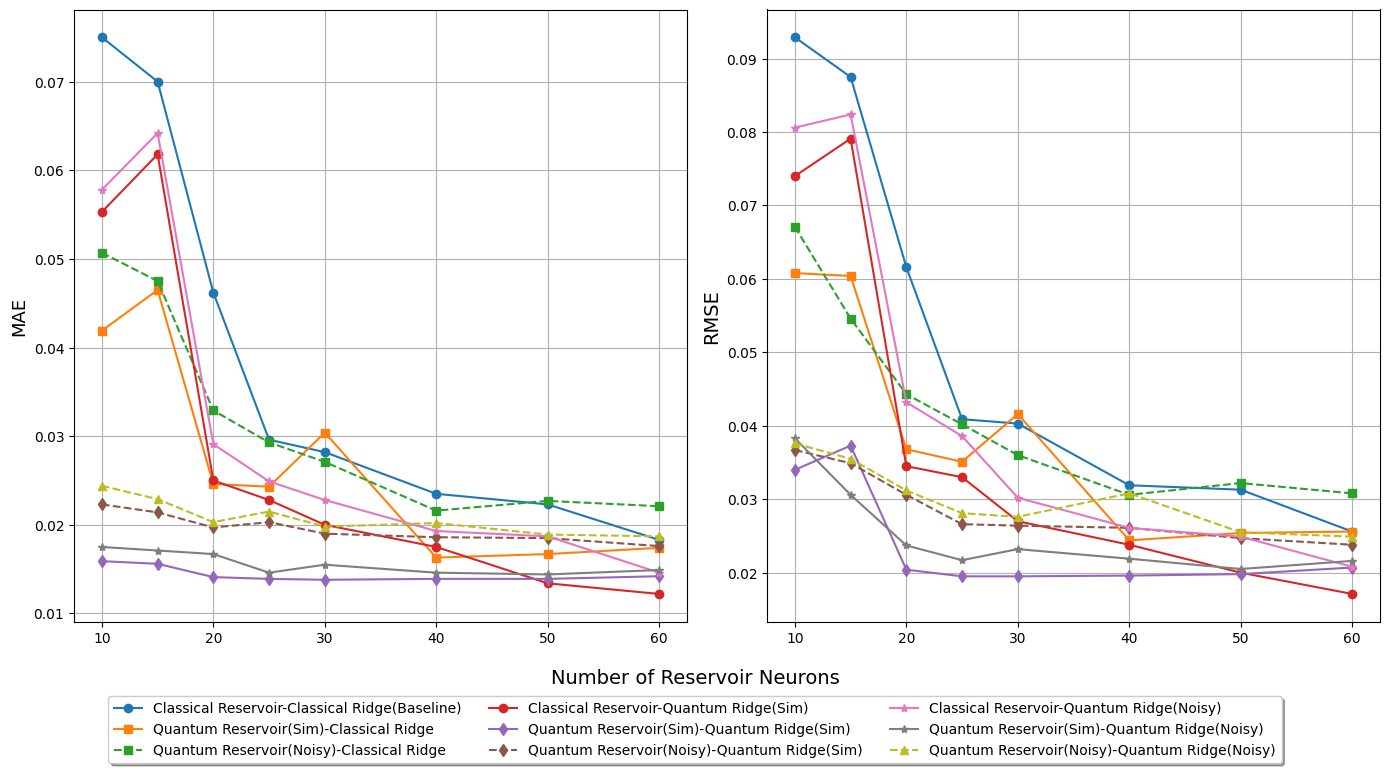}
\caption{Mean Absolute Error and Root Mean Square Error distribution for all Architecture 3 (SRMR) models considered.}
\label{fig6}
\end{figure*}

\begin{figure*}[!h]
\centering
\includegraphics[width=0.7\textwidth]{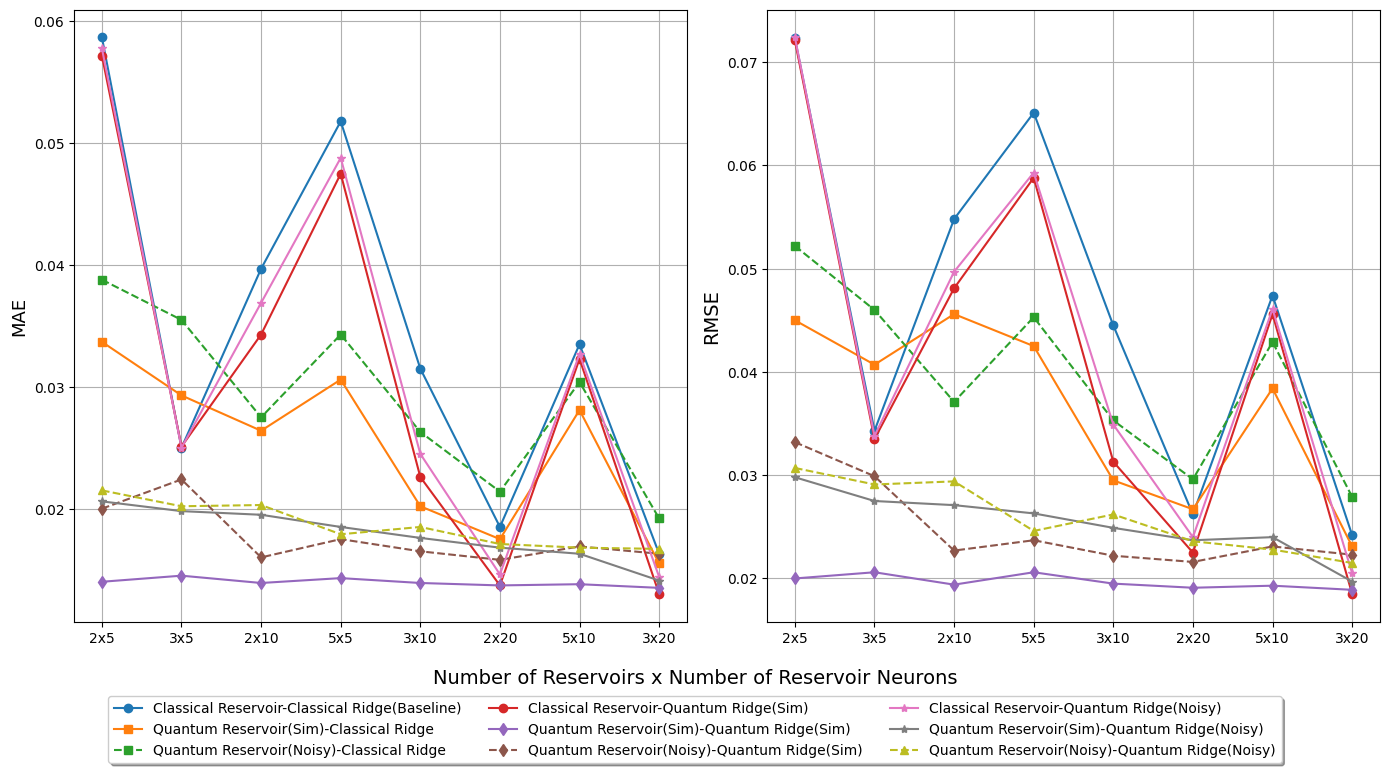}
\caption{Mean Absolute Error and Root Mean Square Error distribution for all Architecture 4 (MRMR) models considered.\vspace{-10pt}}
\label{fig7}
\end{figure*}

\begin{table}[h]
\centering
\caption{Architecture 3 {(SRMR)} best-performing model in accordance with both the number of reservoir neurons and
the number of ridge instances.}
\label{table:tabArc3tot}
\addtolength{\tabcolsep}{-0.26em}
\resizebox{\columnwidth}{!}{%
\begin{tabular}{c|c|c|c|ccc}
\hline
\textbf{Number} & \textbf{Ridge} & \textbf{Reservoir Type} & \textbf{Ridge Type}
 & MAE & RMSE & $R^2$ \\
 \textbf{of Neurons} & \textbf{Instances} & & & & &\\
\hline
10  & 2 & Quantum (Sim.) & Quantum (Noisy.) & 0.0159 & 0.0340 & 0.9557\\
20 &  2 & Quantum (Sim.) & Quantum (Sim.) & 0.0141 & 0.0204 & 0.9841\\
40 & 2 & Classical (Sim.) & Quantum (Sim.)  & 0.0139 & 0.0196 & 0.9853\\
15 & 3 & Quantum (Noisy.) & Quantum (Sim.) & 0.0214 & 0.0349 & 0.9530\\
30 & 3 & Quantum (Sim.) & Quantum (Sim.) & 0.0138 & 0.0195 & 0.9855 \\
60 & 3 &  \textbf{Classical} & \textbf{Quantum (Sim.)}  & \textbf{0.0122} & \textbf{0.0171} & \textbf{0.9888} \\
25 & 5 & Quantum (Sim.) & Quantum (Sim.)  & 0.0139 & 0.0195 & 0.9854 \\
50 & 5 & Quantum (Sim.) & Quantum (Sim.)  &  0.0139 & 0.0198 & 0.9851  \\
\hline
\end{tabular}
}
\end{table}

Similarly to all previous Architectures, the use of a quantum instead of a classical layer as the output of our models assists them to perform better. However, the difference between the classical and quantum ridge output is more obvious in this case, as depicted in Fig.~\ref{fig6}. {In fact, the impact of the distributed quantum kernels can be verified by the comparison to the baseline, as this Architecture yields an average reduction of 54.2\% in MAE across all cases and 50.4\% in RMSE, which are two times ($2 \times$) better compared to the SRSR and MRSR configurations that were previously explored in Subsections \ref{sec:subsec321} and \ref{sec:subsec322}.} Moreover, in this Architecture, there is an observable deviation in performance between the quantum and the classical reservoir models, especially when the number of neurons is relatively small. More specifically, quantum reservoir models are capable of performing about three times better with 10-20 neurons than their classical counterparts. {For this SRMR configuration, the best overall performance improvement compared to its baseline is encountered for the case of 10 neurons and two ridge instances (smallest tested network), with a 78.8\% reduction in MAE and 63.4\% in RMSE.} Last but not least, a noticeable aspect of the fully quantum models that fall in this Architecture is the small performance variance between models with different numbers of neurons, in contrast to all other models, which exhibit significant fluctuations in their performance in relation to the number of neurons.

\subsubsection{Architecture 4: Multi-Reservoir Multi-Readout (MRMR)}
\label{sec:subsec324}
{Finally, combining the advantages of both reservoir and readout distribution, an additional Architecture is introduced, representing a fully modular pipeline designed for maximum parallelization, as both layers are partitioned across a distributed multi-QPU environment.} Thus, in this last Architecture, all models comprise multiple reservoirs and ridge output instances, as presented in Fig. (\ref{figA1b}b). The number of ridge instances is equal to the number of reservoirs, so each reservoir is connected to a single ridge, and together they form a fully distributed system. Table \ref{table:tabArc4tot} presents only the best overall performing models with respect to both the combination of the number of reservoir neurons and the number of neurons in each reservoir (``Res-Num $\times$ Res-Neur") and the combination of the number of ridge instances. As before, the detailed results of all models' executions are provided in Tables \ref{table:tabC10} - \ref{table:tabC12} in Appendix \ref{appC}.



\begin{table}[h]
\centering
\caption{Architecture 4 {(MRMR)} best-performing models in accordance with the number of reservoirs and reservoir neurons and
the number of ridge instances}
\label{table:tabArc4tot}
\addtolength{\tabcolsep}{-0.26em}
\resizebox{\columnwidth}{!}{%
\begin{tabular}{c|c|c|c|ccc}
\hline
\textbf{Res-Num x} & \textbf{Ridge} & \textbf{Reservoir Type} & \textbf{Ridge Type}
 & MAE & RMSE & $R^2$ \\
 \textbf{Res-Neur} & \textbf{Instances} & & & & &\\
\hline
2x5  & 2 & Quantum (Sim.) & Quantum (Sim.) & 0.014 & 0.0200 & 0.9847\\
2x10 &  2 & Quantum (Sim.) & Quantum (Sim.) & 0.0139 & 0.0194 & 0.9856\\
2x20 & 2 & Quantum (Sim.) & Quantum (Sim.)  & 0.0137 & 0.0191 & 0.9858\\
3x5 & 3 & Quantum (Sim.) & Quantum (Sim.) & 0.0145 & 0.0206 & 0.9837\\
3x10 & 3 & Quantum (Sim.) & Quantum (Sim.) & 0.0139 & 0.0195 & 0.9855 \\
3x20 & 3 &  \textbf{Classical} & \textbf{Quantum (Sim.)}  & \textbf{0.0130} & \textbf{0.0185} & \textbf{0.9869} \\
5x5 & 5 & Quantum (Sim.) & Quantum (Sim.)  & 0.0143 & 0.0206 & 0.9838 \\
5x10 & 5 & Quantum (Sim.) & Quantum (Sim.)  &  0.0138 & 0.0193 & 0.9857  \\
\hline
\end{tabular}
}
\end{table}

As with the preceding Architectures, the forecasting performance presented in Table \ref{table:tabArc4tot} reveals the superiority of the quantum ridge layer when used as the output layer of our models. Additionally, the performance of the noisy models in this Architecture is the best overall, across all reservoir layer variations, compared to the noisy models created in previous Architectures. {The fully modular MRMR configuration (Architecture 4) achieves balanced performance gains, with a sustained average improvement of 52.1\% in MAE and 51.4\% in RMSE across all cases and a peak performance improvement in the case of the smallest investigated network (2 reservoirs with 5 neurons each and 2 ridge instances). Those results are similar to those of the SRMR Architecture that was previously examined in Subsection~\ref{sec:subsec323}, highlighting the significance of the output layer in the overall performance of the network and showcasing clear benefits in using quantum output layers, even more in distributed arrangements.} Moreover, in Fig.~\ref{fig7} it is evident that all classical and hybrid models, consisting of a classical reservoir or a classical ridge layer, present significant fluctuations in their performance in relation to the combination of the number of reservoirs and the number of neurons in each reservoir. However, fully quantum models demonstrate a more stable and overall better performance across all numbers of reservoirs and reservoir neurons. This is something that was also apparent in some Architecture 2 (MRSR) models, where only the reservoir layer consisted of multiple subcomponents, indicating that fully quantum models are able to achieve more consistent performance with distributed structures than hybrid or fully classical ones.

\subsubsection{Cross-Architecture Comparison}

In light of the above, we provide Table~\ref{table:tab12} where we present the best Architectures for each total number of neurons in the reservoir layer and their corresponding forecasting performance. It is apparent that, for a small number of neurons, fully quantum models, perform better than their hybrid counterparts. Furthermore, we observe that the quantum ridge output outperforms its classical counterpart in all cases. Finally, the distributed models, either fully (Architecture 4 - MRMR) or partially (Architectures 2 and 3 - MRSR and SRMR respectively) tend to perform better compared to the single-node ones (Architecture 1 - SRSR). {This is directly reflected in Table~\ref{table:tab12}, where for all different cases with different total numbers of neurons, the best-performing configuration involves a level of distribution either in the reservoir or in the output layer (or even in both).} Thus, by distributing model sub-components across multiple QPUs, we can scale under per-device qubit constraints while achieving {comparable} forecasting performance. {This reduction follows a linear trend, where the factor of reduction of the qubit requirement is equal to the number of the sub-reservoirs or the output layer subcomponents.}

{Apart from that, the executed noisy simulations that are mapped to noise data derived from actual hardware, as described in Section~\ref{subesec:performance} and analyzed in Appendix Table~\ref{table:tabAppA} and Table~\ref{table:tabAppB}, provided some additional useful insight on the operation of distributed quantum Architectures. More specifically, when comparing fully quantum models that include both quantum reservoirs and output layers, the MRSR Architecture, which distributes the quantum reservoir, consistently outperforms the SRSR Architecture across all investigated neuron counts, as can be seen by a direct comparison between Appendix Table~\ref{table:tabC3} and Table~\ref{table:tabC6}. Additionally, the fully distributed MRMR Architecture provides lower error metrics compared to the single node SRSR configurations for larger network sizes, from 40 neurons and up, as seen by comparing Appendix Table~\ref{table:tabC3} with Table~\ref{table:tabC12}. On the contrary, a similar behavior is not observed in the case of distributing only the ridge output layer, as the comparison of Appendix Table~\ref{table:tabC3}, related to the SRSR Architecture, with Table~\ref{table:tabC8}, related to the SRMR Architecture, clearly indicates.}
 
{This behavior indicates that the distribution of quantum components, and mainly the quantum reservoir, across multiple QPUs can also act as a potential noise mitigation strategy. By partitioning the quantum components into smaller parallel circuits, the distributed Architectures suffer less from noise accumulation, which is inherently present in larger unified quantum circuits. This behavior and difference between distributed and non-distributed configurations is expected to be even more intense when the employed circuits are larger and of greater depth.}

\begin{table}[h]
\centering
\caption{Best-performing models in accordance with the total number of reservoir neurons.}
\label{table:tab12}
\addtolength{\tabcolsep}{-0.26em}
\resizebox{\columnwidth}{!}{%
\begin{tabular}{c|c|c|c|ccc}
\hline
\textbf{Number} & \textbf{Arch.} & \textbf{Reservoir Type} & \textbf{Ridge Type}
 & MAE & RMSE & $R^2$ \\
 \textbf{of Neurons} & \textbf{Category} & & & & &\\
\hline
10 & MRSR  & Distributed-Quantum & Single-Quantum & 0.0139 & 0.0196 & 0.9853\\
20 & MRMR &  Distributed-Quantum & Distributed-Quantum & 0.0139 & 0.0194 & 0.9856\\
40 & MRMR &  Distributed-Quantum & Distributed-Quantum & 0.0137 & 0.0191 & 0.9858\\
15 & MRSR & Distributed-Quantum & Single-Quantum & 0.0142 & 0.0199 & 0.9848\\
30 & SRMR &  Single-Quantum & Distributed-Quantum & 0.0138 & 0.0195 & 0.9855\\
60 & MRSR &  \textbf{Distributed-Classical} & \textbf{Single-Quantum} & \textbf{0.0113} & \textbf{0.0158} & \textbf{0.9904}\\
25 & SRMR &  Single-Quantum & Distributed-Quantum & 0.0139 & 0.0195 & 0.9854 \\
50 & MRMR &  Distributed-Quantum & Distributed-Quantum & 0.0138 & 0.0193 & 0.9857 \\
\hline
\end{tabular}
}
\end{table}

\section{Conclusions}
\label{sec:conclusions}
In this work, we introduced and evaluated numerous hybrid and pure quantum reservoir computing models against a power load forecasting task. These models consist of a reservoir and a ridge output layer with one or more sub-components comprising these layers. More specifically, we opted for distributed individual components, which led to four (4) different network Architectures; three of them comprise either partially or fully distributed models, whereas the other one consists of models with a single-node on each of their layers. For our simulations, we used both ideal and noise simulators provided by IBM and IONQ, through the Qiskit and NVIDIA CUDA-Q frameworks.

Our numerical simulations show that the use of either pure quantum models or hybrid ones, i.e. by substituting either the classical reservoir or ridge output layer, with a quantum counterpart, is more beneficial than using classical RC models. Furthermore, pure quantum models, either fully or partially distributed, achieve consistently higher performance than hybrid ones, when the reservoir layer consists of few neurons ($<30$ in our tested neuron counts). {Thus, these quantum-enhanced configurations demonstrate an inherent robustness and stability to scaling, maintaining high precision even at reduced dimensionality}. The best overall performance was achieved by a 60 neuron hybrid RC model comprising three 20-neuron reservoirs and a single-node quantum ridge output. All these results demonstrate the significance of using quantum components, either in combination with classical ones, forming hybrid models, or even standalone, for forecasting purposes. {This outcome is further reinforced by the comparison with classical baselines, where across all four (4) different Architectures and all different configurations for each Architecture, the improvement in performance when using fully or partially quantum networks reaches up to 78.8\% and 72.3\% in terms of reducing MAE and RMSE, respectively. The cases of maximum improvement that are mentioned here are both encountered for configurations that use distributed quantum ridge output layers, highlighting their merit compared to classical distributed outputs.}

The presented models are scalable and compatible with existing limitations of the NISQ technology. Additionally, our distributed Architectures, based on their inherent modularity and hardware-agnostic nature, enable scaling under qubit limits and heterogeneity, making their deployment across different quantum hardware a straightforward process and aligning with current approaches for hybrid HPC-QC integration and multi-QPU orchestration \cite{rallis2025interfacing}, without sacrificing performance.

Looking ahead, several avenues for future research emerge from this study. In particular, one may test different quantum circuits in all models' layers and compare their efficacy against the current state of the art quantum RC models. Moreover, extending these distributed RC models to larger-scale ones and testing them on different quantum hardware constitute important aspects for further study. This study facilitates the flexibility of the proposed distributed Architecture both in terms of sub-component number, as well as the hardware agnosticism. As an example, a different number of reservoirs and ridge instances in the output layer can be considered, where different QPUs may be utilized for each of the aforementioned sub-components. 
Together, all of these developments are anticipated to contribute to the improved applicability and broader impact of the proposed methodology.

\appendices

\section{}
\label{appA}
\renewcommand{\thefigure}{A\arabic{figure}}
\setcounter{figure}{0}
This section presents the structural diagrams of architectures 1 {(SRSR) and 4 (MRMR)}, as presented in Section~\ref{sec:architectures}. As a reminder, SRSR refers to models comprising a single reservoir and ridge output layer, whereas MRMR refers to models comprising multiple reservoirs and ridge output instances.


\begin{figure}[h]
\centering
\includegraphics[width=\columnwidth]{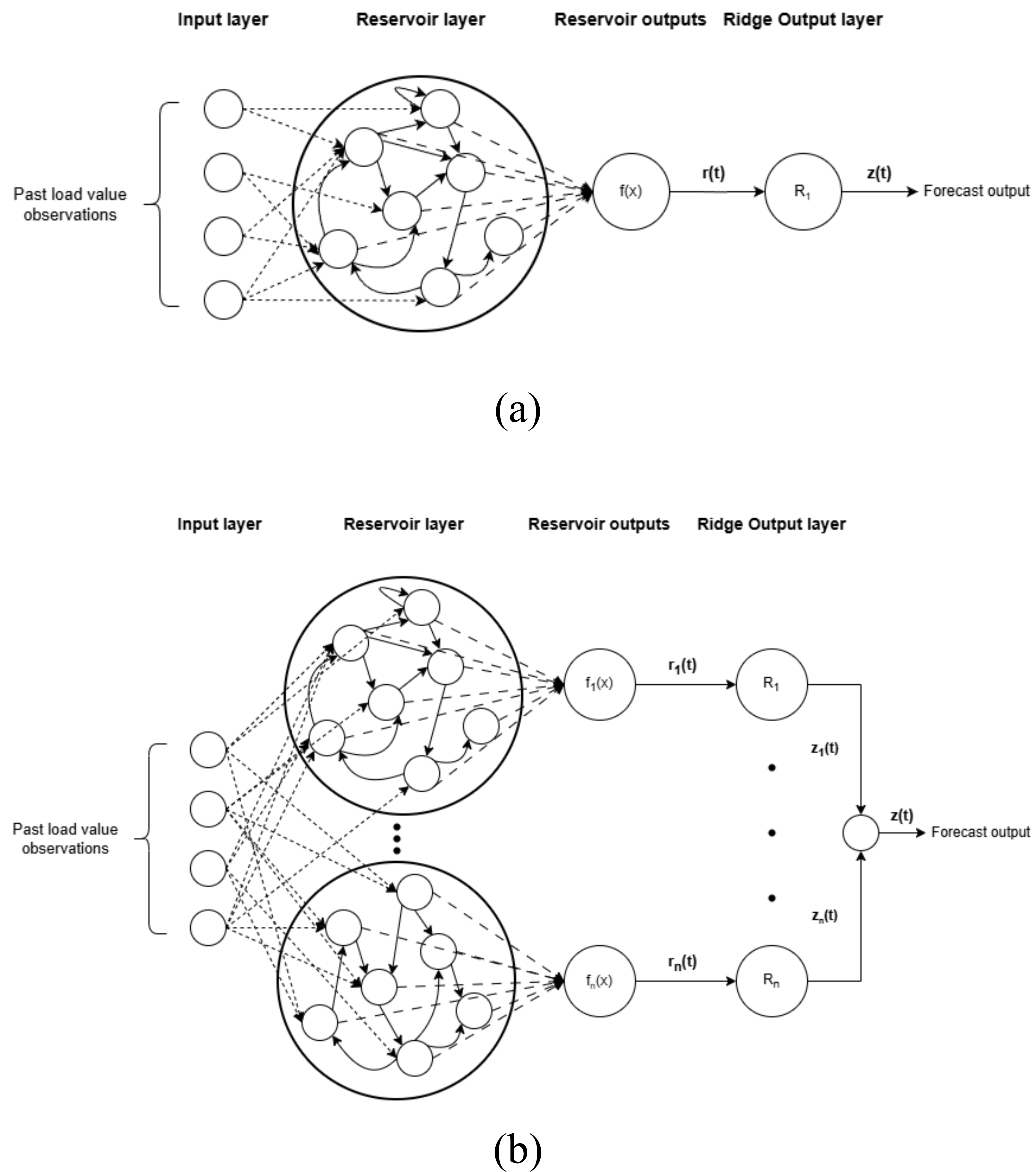}
\caption{(a) Architecture 1 (SRSR) and (b) Architecture 4 (MRMR) as introduced in Table \ref{table:tab1}. In both figures $f_i(x)$ or $f(x)$ represents the concatenation of all $i_{th}$ reservoir data into an output vector, without the application of an activation function.}
\label{figA1}
\phantomsection\label{figA1a}
\phantomsection\label{figA1b}
\end{figure}

\section{}{\label{appB}}
\renewcommand{\thetable}{B\arabic{table}}
\setcounter{table}{0}
In this section, we include the calibration data for `IBM Brisbane', `IBM Marrakesh', and `IONQ Aria - 1' QPUs at the time when our experiments took place. They can be seen in Table~\ref{table:tabAppA} and Table~\ref{table:tabAppB}.

\begin{table}[h]
\centering
\caption{Calibration data for `IBM Brisbane' and `IBM Marrakesh'.}
\label{table:tabAppA}
\resizebox{\columnwidth}{!}{%
\begin{tabular}{c|c|c}
\hline
 & \multicolumn{1}{c|}{\textbf{IBM Brisbane}} 
 & \multicolumn{1}{c}{\textbf{IBM Marrakesh}} \\
 \hline
 \textbf{Processor Type} & \multicolumn{1}{c|}{Eagle r3} & \multicolumn{1}{c}{Heron r2} \\
 \hline
 \textbf{Specifications} & \multicolumn{2}{c}{\textbf{Value}} \\
\hline
2Q error (best)	& 3.28e-3 & 1.26e-3\\
2Q error (layered) & 2.07e-2 & 7.53e-3\\
Median ECR error & 7.519e-3 & - \\
Median CZ error & - &  3.351e-3\\
Median SX error	& 2.236e-4 & 2.304e-4\\
Median readout error &	1.660e-2 & 1.038e-2\\
Median T1 & 230.85 us & 219.88 us\\
Median T2 & 154.59 us & 118.27 us\\
\hline
\end{tabular}
}
\end{table}

\begin{table}[h]
\centering
\caption{Calibration data for `IONQ Aria-1'.}
\label{table:tabAppB}
\begin{tabular}{c|c}
\hline
 \multicolumn{2}{c}{\textbf{IONQ Aria-1}}\\
 \hline
 \textbf{Specifications} & \multicolumn{1}{c}{\textbf{Value}} \\
\hline
Measurement fidelity (median) & 0.9951\\
1Q fidelity (median) & 0.9998\\
2Q fidelity (median) & 0.9858\\
Median T1 & 100 s\\
Median T2 & 1 s\\
\hline
\end{tabular}
\end{table}

\section{}\label{appC}
In this section, all detailed forecasting performance tables are demonstrated. For each architecture, we present three tables; the first one refers to the models performance when the ridge output layer is classically implemented, whereas the later two tables refer to the models' performance when the ridge output layer is quantum-kernel-based, executed on both the ideal and on the noisy simulator.

\renewcommand{\thesubsection}{C.\arabic{subsection}}
\subsection{Architecture 1: Single Reservoir Single Readout (SRSR)}
Models that fall into this category consist of a single reservoir and a single ridge output layer. In Table \ref{table:tabC1} the forecasting performance of each model is demonstrated when the output layer consists of a classical ridge instance:

\renewcommand{\thetable}{C\arabic{table}}
\setcounter{table}{0}

\begin{table}[h]
\caption{Architecture 1 (SRSR) model performance with classical Ridge output.}
\label{table:tabC1}
\addtolength{\tabcolsep}{-0.13em}
\resizebox{\columnwidth}{!}{%
\begin{tabular}{c|ccc|ccc|ccc}
\hline
 & \multicolumn{3}{c|}{\textbf{Classical Reservoir}} 
 & \multicolumn{6}{c}{\textbf{Quantum Reservoir}} \\
\hline
 & \multicolumn{3}{c|}{Simulation}
 & \multicolumn{3}{c|}{Simulation}
 & \multicolumn{3}{c}{Noisy} \\
\hline
\textbf{Reservoir}
 & MAE & RMSE & $R^2$
 & MAE & RMSE & $R^2$
 & MAE & RMSE & $R^2$ \\
 \textbf{Neurons} & & & & & & & & & \\
\hline
10 & 0.0458 & 0.0627 & 0.8494 & 0.0230 & 0.0340 & 0.9557 & 0.0257 & 0.0363 & 0.9495 \\
20 & 0.0225 & 0.0315 & 0.9619 & 0.0180 & 0.0262 & 0.9738 & 0.0245 & 0.0327 & 0.9591 \\
40 & 0.0151 & 0.0222 & 0.9812 & 0.0182 & 0.0287 & 0.9714 & 0.0199 & 0.0281 & 0.9698 \\
15 & 0.0455 & 0.0599 & 0.8629 & 0.0189 & 0.0269 & 0.9722 & 0.0321 & 0.0469 & 0.9158 \\
30 & 0.0159 & 0.0234 & 0.9789 & 0.0177 & 0.0284 & 0.9692 & 0.0223 & 0.0308 & 0.9637 \\
60 & \textbf{0.0136} & \textbf{0.0201} & \textbf{0.9845} & 0.0162 & 0.0240 & 0.9779 & 0.0175 & 0.0264 & 0.9726 \\
25 & 0.0174 & 0.0257 & 0.9747 & 0.0180 & 0.0380 & 0.9647 & 0.0236 & 0.0315 & 0.9605 \\
50 & 0.0142 & 0.0205 & 0.9839 & 0.0172 & 0.0265 & 0.9730 & 0.0185 & 0.0277 & 0.9713 \\
\hline
\end{tabular}
}
\end{table}
Similarly, in Tables \ref{table:tabC2} and \ref{table:tabC3} the forecasting performance of each model is demonstrated when the output layer consists of a quantum-kernel-based ridge instance, executed on ideal and on noisy simulator, accordingly:

\begin{table}[h]
\caption{Architecture 1 (SRSR) model performance with quantum Ridge output, executed on \textbf{ideal} simulator.}
\label{table:tabC2}
\addtolength{\tabcolsep}{-0.13em}
\resizebox{\columnwidth}{!}{%
\begin{tabular}{c|ccc|ccc|ccc}
\hline
 & \multicolumn{3}{c|}{\textbf{Classical Reservoir}} 
 & \multicolumn{6}{c}{\textbf{Quantum Reservoir}} \\
\hline
 & \multicolumn{3}{c|}{Simulation}
 & \multicolumn{3}{c|}{Simulation}
 & \multicolumn{3}{c}{Noisy} \\
\hline
\textbf{Reservoir}
 & MAE & RMSE & $R^2$
 & MAE & RMSE & $R^2$
 & MAE & RMSE & $R^2$ \\
 \textbf{Neurons} & & & & & & & & & \\
\hline
10 & 0.0389 & 0.0552 & 0.8834 & 0.0141 & 0.0198 & 0.9850 & 0.0207 & 0.0277 & 0.9707 \\
20 & 0.0164 & 0.0233 & 0.9792 & 0.0141 & 0.0201 & 0.9845 & 0.0191 & 0.0258 & 0.9745 \\
40 & 0.0141 & 0.0204 & 0.9841 & 0.0142 & 0.0205 & 0.9839 & 0.0206 & 0.0278 & 0.9705 \\
15 & 0.0269 & 0.0365 & 0.9490 & 0.0141 & 0.0204 & 0.9841 & 0.0209 & 0.0289 & 0.9680 \\
30 & 0.0142 & 0.0195 & 0.9854 & 0.0140 & 0.0199 & 0.9848 & 0.0200 & 0.0276 & 0.9708 \\
60 & \textbf{0.0116} & \textbf{0.0165} & \textbf{0.9896} & 0.0146 & 0.0220 & 0.9815 & 0.0174 & 0.0238 & 0.9766 \\
25 & 0.0158 & 0.0220 & 0.9815 & 0.0139 & 0.0197 & 0.9851 & 0.0193 & 0.0264 & 0.9731 \\
50 & 0.0149 & 0.0292 & 0.9674 & 0.0141 & 0.0202 & 0.9840 & 0.0188 & 0.0251 & 0.9753 \\
\hline
\end{tabular}
}
\end{table}

\begin{table}[h]
\caption{Architecture 1 (SRSR) model performance with quantum Ridge output, executed on \textbf{noisy} simulator.}
\label{table:tabC3}
\addtolength{\tabcolsep}{-0.13em}
\resizebox{\columnwidth}{!}{%
\begin{tabular}{c|ccc|ccc|ccc}
\hline
 & \multicolumn{3}{c|}{\textbf{Classical Reservoir}} 
 & \multicolumn{6}{c}{\textbf{Quantum Reservoir}} \\
\hline
 & \multicolumn{3}{c|}{Simulation}
 & \multicolumn{3}{c|}{Simulation}
 & \multicolumn{3}{c}{Noisy} \\
\hline
\textbf{Reservoir}
 & MAE & RMSE & $R^2$
 & MAE & RMSE & $R^2$
 & MAE & RMSE & $R^2$ \\
 \textbf{Neurons} & & & & & & & & & \\
\hline
10 & 0.0411 & 0.0583 & 0.8712 &  0.0178 &  0.0275 & 0.9695 & 0.0214 & 0.0285 & 0.9698 \\
20 & 0.0214 & 0.0297 & 0.9632 &  0.0166 &  0.0261 & 0.9734 & 0.0196 & 0.0267 & 0.9731 \\
40 & 0.0146 & 0.0214 & 0.9827 &  0.0153 & 0.0242 & 0.9763 & 0.0189 & 0.0258 & 0.9747 \\
15 & 0.0386 & 0.0528 & 0.8865 &  0.0172 &  0.0264 & 0.9728 & 0.0208 & 0.0282 & 0.9702 \\
30 & 0.015 & 0.0216 & 0.9814 &  0.0157 & 0.0248 & 0.9757 & 0.0193 & 0.0261 & 0.9736 \\
60 & \textbf{0.0129} & \textbf{0.0188} & \textbf{0.9875} & 0.0148 & 0.0219 & 0.9823 & 0.0176 & 0.0241 & 0.9762 \\
25 & 0.0171 & 0.0245 & 0.9763 &  0.0164 & 0.0253 & 0.9754 & 0.0195 & 0.0262 & 0.9738 \\
50 & 0.0144 & 0.0262 & 0.9691 &  0.0152 & 0.0228 & 0.9807 & 0.0185 & 0.0249 & 0.9755 \\
\hline
\end{tabular}
}
\end{table}

\subsection{Architecture 2: Multi-Reservoir Single-Readout (MRSR)}
Models that fall into this category consist of multiple reservoirs and a single ridge output layer. In Table \ref{table:tabC4} the forecasting performance of each model is demonstrated when the output layer consists of a classical ridge instance. In all tables, the abbreviation `Res-Num $\times$ Res-Neur' indicates the number of reservoirs in the reservoir layer and the number of neurons comprising each reservoir, accordingly.

\begin{table}[h!]
\caption{Architecture 2 (MRSR) model performance with classical Ridge output.}
\label{table:tabC4}
\addtolength{\tabcolsep}{-0.14em}
\resizebox{\columnwidth}{!}{%
\begin{tabular}{c|ccc|ccc|ccc}
\hline
 & \multicolumn{3}{c|}{\textbf{Classical Reservoir}} 
 & \multicolumn{6}{c}{\textbf{Quantum Reservoir}} \\
\hline
 & \multicolumn{3}{c|}{Simulation}
 & \multicolumn{3}{c|}{Simulation}
 & \multicolumn{3}{c}{Noisy} \\
\hline
\textbf{Res-Num x}
 & MAE & RMSE & $R^2$
 & MAE & RMSE & $R^2$
 & MAE & RMSE & $R^2$ \\
 \textbf{Res-Neur} & & & & & & & & & \\
\hline
2x5 & 0.0212 & 0.0323 & 0.9600 & 0.0229 & 0.0446 & 0.9239 & 0.0281 & 0.0369 & 0.9479 \\
2x10 & 0.0382 & 0.0530 & 0.8924 & 0.0171 & 0.0248 & 0.9764 & 0.0258 & 0.0350 & 0.9532 \\
2x20 & 0.0175 & 0.0244 & 0.9772 & 0.0168 & 0.0257 & 0.9747 & 0.0198 & 0.0278 & 0.9704 \\
3x5 & 0.0163 & 0.0252 & 0.9756 & 0.0193 & 0.0279 & 0.9702 & 0.0239 & 0.0330 & 0.9585 \\
3x10 & 0.0190 & 0.0296 & 0.9664 & 0.0174 & 0.0344 & 0.9547 & 0.0205 & 0.0285 & 0.9690 \\
3x20 & \textbf{0.0128} & \textbf{0.0189} & \textbf{0.9863} & 0.0159 & 0.0245 & 0.9770 & 0.0179 & 0.0261 & 0.9755 \\
5x5 & 0.0174 & 0.0274 & 0.9713 & 0.0178 & 0.0292 & 0.9673 & 0.0246 & 0.0335 & 0.9578 \\
5x10 & 0.0175 & 0.0285 & 0.9690 & 0.0171 & 0.0252 & 0.9758 & 0.0183 & 0.0268 & 0.9739 \\
\hline
\end{tabular}
}
\end{table}
Similarly, the forecasting capability of all hybrid or pure quantum models consisting of multiple reservoirs and a quantum ridge output is presented in Tables \ref{table:tabC5} and \ref{table:tabC6}. In Table \ref{table:tabC5} the output layer of all models was executed on an ideal simulator, whereas in Table \ref{table:tabC6} the  output layer of all models was executed on a noisy one (`IBM Brisbane').

\begin{table}[h!]
\caption{Architecture 2 (MRSR) model performance with quantum Ridge output, executed on \textbf{ideal} simulator.}
\label{table:tabC5}
\addtolength{\tabcolsep}{-0.14em}
\resizebox{\columnwidth}{!}{%
\begin{tabular}{c|ccc|ccc|ccc}
\hline
 & \multicolumn{3}{c|}{\textbf{Classical Reservoir}} 
 & \multicolumn{6}{c}{\textbf{Quantum Reservoir}} \\
\hline
 & \multicolumn{3}{c|}{Simulation}
 & \multicolumn{3}{c|}{Simulation}
 & \multicolumn{3}{c}{Noisy} \\
\hline
\textbf{Res-Num x}
 & MAE & RMSE & $R^2$
 & MAE & RMSE & $R^2$
 & MAE & RMSE & $R^2$ \\
 \textbf{Res-Neur} & & & & & & & & & \\
\hline
2x5  & 0.0165 & 0.0236 & 0.9787 & 0.0139 & 0.0196 & 0.9853 & 0.0204 & 0.0289 & 0.9681 \\
2x10 & 0.0343 & 0.0479 & 0.9122 & 0.0140 & 0.0198 & 0.9850 & 0.0173 & 0.0244 & 0.9773 \\
2x20 & 0.0142 & 0.0317 & 0.9615 & 0.0140 & 0.0202 & 0.9845 & 0.0158 & 0.0225 & 0.9807 \\
3x5  & 0.0157 & 0.0223 & 0.9810 & 0.0142 & 0.0199 & 0.9848 & 0.0169 & 0.0246 & 0.9769 \\
3x10 & 0.0147 & 0.0206 & 0.9837 & 0.0141 & 0.0204 & 0.9841 & 0.0176 & 0.0243 & 0.9774 \\
3x20 & \textbf{0.0113} & \textbf{0.0158} & \textbf{0.9904} & 0.0145 & 0.0213 & 0.9826 & 0.0147 & 0.0209 & 0.9833 \\
5x5  & 0.0146 & 0.0207 & 0.9837 & 0.0142 & 0.0202 & 0.9844 & 0.0165 & 0.0241 & 0.9791 \\
5x10 & 0.0148 & 0.0224 & 0.9807 & 0.0143 & 0.0211 & 0.9830 & 0.0152 & 0.0217 & 0.9816 \\
\hline
\end{tabular}
}
\end{table}

\begin{table}[h!]
\caption{Architecture 2 (MRSR) model performance with quantum Ridge output, executed on \textbf{noisy} simulator.}
\label{table:tabC6}
\addtolength{\tabcolsep}{-0.14em}
\resizebox{\columnwidth}{!}{%
\begin{tabular}{c|ccc|ccc|ccc}
\hline
 & \multicolumn{3}{c|}{\textbf{Classical Reservoir}} 
 & \multicolumn{6}{c}{\textbf{Quantum Reservoir}} \\
\hline
 & \multicolumn{3}{c|}{Simulation}
 & \multicolumn{3}{c|}{Simulation}
 & \multicolumn{3}{c}{Noisy} \\
\hline
\textbf{Res-Num x}
 & MAE & RMSE & $R^2$
 & MAE & RMSE & $R^2$
 & MAE & RMSE & $R^2$ \\
 \textbf{Res-Neur} & & & & & & & & & \\
\hline
2x5  & 0.0183 & 0.0274 & 0.9735 & 0.0176 & 0.0256 & 0.9748 & 0.0211 & 0.0296 & 0.9674 \\
2x10 & 0.0335 & 0.0458 & 0.9216 & 0.0158 & 0.0217 & 0.9813 & 0.0188 & 0.0263 & 0.9726  \\
2x20 & 0.015 & 0.0279 & 0.9603 & 0.0153 & 0.0205 & 0.9828 & 0.0159 & 0.0233 & 0.9801  \\
3x5  & 0.0176 & 0.0245 & 0.9787 & 0.0169 & 0.0231 & 0.9791 & 0.0192 & 0.0269 & 0.9718 \\
3x10 & 0.0168 & 0.0239 & 0.9792 & 0.0155 & 0.0207 & 0.983 & 0.0172 & 0.0238 & 0.9789  \\
3x20 & 0.0123 & 0.0163 & 0.9883 & 0.0145 & 0.0203 & 0.9842 & 0.0148 & 0.0207 & 0.9825  \\
5x5  & 0.0152 & 0.0224 & 0.9815 & 0.0151 & 0.0204 & 0.9846 & 0.0153 & 0.0219 & 0.9808  \\
5x10 & 0.0149 & 0.0203 & 0.9809 & 0.0147 & 0.0196 & 0.9854 & 0.015 & 0.0212 & 0.9811  \\
\hline
\end{tabular}
}
\end{table}

\subsection{Architecture 3: Single-Reservoir Multi-Readout (SRMR)}
In this Architecture, all models consist of a single reservoir and multiple ridge output instances, classical or quantum. In Tables \ref{table:tabC7}, \ref{table:tabC8} and \ref{table:tabC9} we present the forecasting results of the models tested, consisting of classical and quantum ridge instances, accordingly. All results demonstrated in Table \ref{table:tabC8} were derived from ideal simulations, while those in Table \ref{table:tabC9} were obtained from noisy ones. In both tables, ``Res Num" refers to the number of neurons in the reservoir layer. In Table \ref{table:tabC9} the abbreviation ``Out-Num $\times$ Q-Out" refers to the number of ridge instances (``Out-Num") and the number of qubits exploited in each quantum feature map (``Q-Out") at each instance.

\begin{table}[h!]
\caption{Architecture 3 (SRMR) model performance with multiple classical Ridge output instances.}
\label{table:tabC7}
\addtolength{\tabcolsep}{-0.26em}
\resizebox{\columnwidth}{!}{%
\begin{tabular}{c|c|ccc|ccc|ccc}
\hline
 & & \multicolumn{3}{c|}{\textbf{Classical Reservoir}} 
 & \multicolumn{6}{c}{\textbf{Quantum Reservoir}} \\
\hline
 & & \multicolumn{3}{c|}{Simulation}
 & \multicolumn{3}{c|}{Simulation}
 & \multicolumn{3}{c}{Noisy} \\
\hline
\textbf{Res} & \textbf{Ridge}
 & MAE & RMSE & $R^2$
 & MAE & RMSE & $R^2$
 & MAE & RMSE & $R^2$ \\
 \textbf{Neur} & \textbf{Instances} & & & & & & & & & \\
\hline
10 & 2 & 0.0750 & 0.0929 & 0.6701 & 0.0419 & 0.0608 & 0.8586 & 0.0507 & 0.0671 & 0.8278 \\
20 & 2 & 0.0462 & 0.0616 & 0.8548 & 0.0246 & 0.0368 & 0.9483 & 0.0329 & 0.0443 & 0.9249 \\
40 & 2 & 0.0235 & 0.0319 & 0.9612 & \textbf{0.0163} & \textbf{0.0244} & \textbf{0.9766} & 0.0216 & 0.0306 & 0.9641 \\
15 & 3 & 0.0700 & 0.0875 & 0.7070 & 0.0465 & 0.0604 & 0.8604 & 0.0475 & 0.0546 & 0.8659 \\
30 & 3 & 0.0282 & 0.0403 & 0.9380 & 0.0304 & 0.0416 & 0.9319 & 0.0271 & 0.0360 & 0.9505 \\
60 & 3 & 0.0183 & 0.0256 & 0.9749 & 0.0174 & 0.0256 & 0.9744 & 0.0221 & 0.0308 & 0.9638 \\
25 & 5 & 0.0296 & 0.0409 & 0.9361 & 0.0243 & 0.0351 & 0.9529 & 0.0293 & 0.0402 & 0.9345 \\
50 & 5 & 0.0223 & 0.0313 & 0.9626 & 0.0167 & 0.0254 & 0.9753 & 0.0227 & 0.0322 & 0.9619 \\
\hline
\end{tabular}
}
\end{table}

\begin{table}[h!]
\caption{Architecture 3 (SRMR) model performance with multiple quantum Ridge output instances, executed on \textbf{ideal} simulator.}
\label{table:tabC8}
\addtolength{\tabcolsep}{-0.26em}
\resizebox{\columnwidth}{!}{%
\begin{tabular}{c|c|ccc|ccc|ccc}
\hline
 & & \multicolumn{3}{c|}{\textbf{Classical Reservoir}} 
 & \multicolumn{6}{c}{\textbf{Quantum Reservoir}} \\
\hline
 & & \multicolumn{3}{c|}{Simulation}
 & \multicolumn{3}{c|}{Simulation}
 & \multicolumn{3}{c}{Noisy} \\
\hline
\textbf{Res} & \textbf{Out-Num}
 & MAE & RMSE & $R^2$
 & MAE & RMSE & $R^2$
 & MAE & RMSE & $R^2$ \\
 \textbf{Neur} & \textbf{x} & & & & & & & & & \\
  & \textbf{Q-Out} & & & & & & & & & \\
\hline
10 & 2x5  & 0.0553 & 0.0740 & 0.7904 & 0.0159 & 0.0340 & 0.9557 & 0.0223 & 0.0367 & 0.9517 \\
20 & 2x10 & 0.0250 & 0.0345 & 0.9544 & 0.0141 & 0.0204 & 0.9841 & 0.0197 & 0.0306 & 0.9713 \\
40 & 2x10 & 0.0175 & 0.0238 & 0.9783 & 0.0139 & 0.0196 & 0.9853 & 0.0186 & 0.0261 & 0.9765 \\
15 & 3x5  & 0.0618 & 0.0791 & 0.7606 & 0.0156 & 0.0373 & 0.9467 & 0.0214 & 0.0349 & 0.9530 \\
30 & 3x10 & 0.0200 & 0.0270 & 0.9721 & 0.0138 & 0.0195 & 0.9855 & 0.0190 & 0.0264 & 0.9758 \\
60 & 3x10 & \textbf{0.0122} & \textbf{0.0171} & \textbf{0.9888} & 0.0142 & 0.0207 & 0.9837 & 0.0176 & 0.0238 & 0.9772 \\
25 & 5x5  & 0.0228 & 0.0330 & 0.9584 & 0.0139 & 0.0195 & 0.9854 & 0.0203 & 0.0266 & 0.9723 \\
50 & 5x10 & 0.0134 & 0.0200 & 0.9847 & 0.0139 & 0.0198 & 0.9851 & 0.0185 & 0.0247 & 0.9761 \\
\hline
\end{tabular}
}
\end{table}

\begin{table}[h!]
\caption{Architecture 3 (SRMR) model performance with multiple quantum Ridge output instances, executed on \textbf{noisy-model} simulators.}
\label{table:tabC9}
\addtolength{\tabcolsep}{-0.26em}
\resizebox{\columnwidth}{!}{%
\begin{tabular}{c|c|ccc|ccc|ccc}
\hline
 & & \multicolumn{3}{c|}{\textbf{Classical Reservoir}} 
 & \multicolumn{6}{c}{\textbf{Quantum Reservoir}} \\
\hline
 & & \multicolumn{3}{c|}{Simulation}
 & \multicolumn{3}{c|}{Simulation}
 & \multicolumn{3}{c}{Noisy} \\
\hline
\textbf{Res} & \textbf{Out-Num}
 & MAE & RMSE & $R^2$
 & MAE & RMSE & $R^2$
 & MAE & RMSE & $R^2$ \\
 \textbf{Neur} & \textbf{x} & & & & & & & & & \\
  & \textbf{Q-Out} & & & & & & & & & \\
\hline
10 & 2x5  & 0.0578 & 0.0806 & 0.7686 & 0.0175 & 0.0383 & 0.9522 & 0.0244 & 0.0376 & 0.9492 \\
20 & 2x10 & 0.0291 & 0.0432 & 0.9235 & 0.0167 & 0.0237 & 0.9788 & 0.0203 & 0.0312 & 0.9685 \\
40 & 2x10 & 0.0193 & 0.0261 & 0.9716 & 0.0146 & 0.0219 & 0.9813 & 0.0202 & 0.0308 & 0.9698 \\
15 & 3x5  & 0.0642 & 0.0824 & 0.7467 & 0.0171 & 0.0336 & 0.9520 & 0.0229 & 0.0355 & 0.9509 \\
30 & 3x10 & 0.0228 & 0.0302 & 0.9609 & 0.0155 & 0.0232 & 0.9796 & 0.0198 & 0.0276 & 0.9716 \\
60 & 3x10 & \textbf{0.0146} & \textbf{0.0208} & \textbf{0.9835} & 0.0149 & 0.0216 & 0.9818 & 0.0187 & 0.0249 & 0.9757 \\
25 & 5x5  & 0.0249 & 0.0386 & 0.9491 & 0.0146 & 0.0217 & 0.9816 & 0.0215 & 0.0281 & 0.9711 \\
50 & 5x10 & 0.0187 & 0.0249 & 0.9752 & 0.0144 & 0.0205 & 0.9827 & 0.0189 & 0.0255 & 0.9745 \\
\hline
\end{tabular}
}
\end{table}

\subsection{Architecture 4: Multi-Reservoir Multi-Readout (MRMR)}
In this last architecture, all models comprise multiple reservoirs and ridge output instances. The forecasting performance of models comprising classical and quantum ridge layers is presented in Tables \ref{table:tabC10}, \ref{table:tabC11}, and \ref{table:tabC12}, accordingly. In all tables, the ``Res Neur" abbreviation refers to the number of reservoirs and the number of neurons in each reservoir, respectively. Similarly, the abbreviation ``Out-Num $\times$ Q-Out" refers to the number of ridge instances (``Out-Num") and the number of qubits exploited in each quantum feature map (``Q-Out") at each instance.

\begin{table}[h!]
\caption{Architecture's 4 {(MRMR)} model performance with multiple classical Ridge output instances.}
\label{table:tabC10}
\addtolength{\tabcolsep}{-0.26em}
\resizebox{\columnwidth}{!}{%
\begin{tabular}{c|c|ccc|ccc|ccc}
\hline
 & & \multicolumn{3}{c|}{\textbf{Classical Reservoir}} 
 & \multicolumn{6}{c}{\textbf{Quantum Reservoir}} \\
\hline
 & & \multicolumn{3}{c|}{Simulation}
 & \multicolumn{3}{c|}{Simulation}
 & \multicolumn{3}{c}{Noisy} \\
\hline
\textbf{Res} & \textbf{Out-Num}
 & MAE & RMSE & $R^2$
 & MAE & RMSE & $R^2$
 & MAE & RMSE & $R^2$ \\
 \textbf{Neur} & \textbf{x} & & & & & & & & & \\
  & \textbf{Q-Out} & & & & & & & & & \\
\hline
2x5  & 2 & 0.0587 & 0.0723 & 0.8002 & 0.0337 & 0.0450 & 0.9225 & 0.0388 & 0.0522 & 0.8957 \\
2x10 & 2 & 0.0397 & 0.0548 & 0.8851 & 0.0264 & 0.0456 & 0.9203 & 0.0275 & 0.0371 & 0.9474 \\
2x20 & 2 & 0.0185 & 0.0262 & 0.9737 & 0.0175 & 0.0267 & 0.9727 & 0.0214 & 0.0296 & 0.9665 \\
3x5  & 3 & 0.0250 & 0.0343 & 0.9551 & 0.0293 & 0.0407 & 0.9304 & 0.0355 & 0.0460 & 0.9189 \\
3x10 & 3 & 0.0315 & 0.0445 & 0.9243 & 0.0202 & 0.0295 & 0.9668 & 0.0263 & 0.0353 & 0.9523 \\
3x20 & 3 & 0.0163 & 0.0242 & 0.9776 & \textbf{0.0155} & \textbf{0.0231} & \textbf{0.9785} & 0.0192 & 0.0279 & 0.9711 \\
5x5  & 5 & 0.0518 & 0.0651 & 0.8377 & 0.0306 & 0.0425 & 0.9310 & 0.0343 & 0.0453 & 0.9203\\
5x10 & 5 & 0.0335 & 0.0474 & 0.9139 & 0.0281 & 0.0384 & 0.9436 & 0.0304 & 0.0429 & 0.9268\\
\hline
\end{tabular}
}
\end{table}

\begin{table}[h!]
\caption{Architecture's 4 {(MRMR)} model performance with multiple quantum Ridge output instances, executed on \textbf{ideal} simulator.}
\label{table:tabC11}
\addtolength{\tabcolsep}{-0.26em}
\resizebox{\columnwidth}{!}{%
\begin{tabular}{c|c|ccc|ccc|ccc}
\hline
 & & \multicolumn{3}{c|}{\textbf{Classical Reservoir}} 
 & \multicolumn{6}{c}{\textbf{Quantum Reservoir}} \\
\hline
 & & \multicolumn{3}{c|}{Simulation}
 & \multicolumn{3}{c|}{Simulation}
 & \multicolumn{3}{c}{Noisy} \\
\hline
\textbf{Res} & \textbf{Out-Num}
 & MAE & RMSE & $R^2$
 & MAE & RMSE & $R^2$
 & MAE & RMSE & $R^2$ \\
 \textbf{Neur} & \textbf{x} & & & & & & & & & \\
  & \textbf{Q-Out} & & & & & & & & & \\
\hline
2x5  & 2x5  & 0.0572 & 0.0721 & 0.8010 & 0.014 & 0.0200 & 0.9847 & 0.0200 & 0.0332 & 0.9579 \\
2x10 & 2x10 & 0.0343 & 0.0481 & 0.9116 & 0.0139 & 0.0194 & 0.9856 & 0.0160 & 0.0227 & 0.9804 \\
2x20 & 2x10 & 0.0137 & 0.0225 & 0.9807 & 0.0137 & 0.0191 & 0.9858 & 0.0158 & 0.0216 & 0.9821 \\
3x5  & 3x5  & 0.0251 & 0.0335 & 0.9572 & 0.0145 & 0.0206 & 0.9837 & 0.0224 & 0.0299 & 0.9657 \\
3x10 & 3x10 & 0.0226 & 0.0313 & 0.9624 & 0.0139 & 0.0195 & 0.9855 & 0.0165 & 0.0222 & 0.9812 \\
3x20 & 3x10 & \textbf{0.0130} & \textbf{0.0185} & \textbf{0.9869} & 0.0135 & 0.0189 & 0.9864 & 0.0163 & 0.0223 & 0.9810 \\
5x5  & 5x5  & 0.0475 & 0.0588 & 0.8675 & 0.0143 & 0.0206 & 0.9838 & 0.0175 & 0.0237 & 0.9798\\
5x10 & 5x10 & 0.0323 & 0.0456 & 0.9205 & 0.0138 & 0.0193 & 0.9857 & 0.0169 & 0.0231 & 0.9804 \\
\hline
\end{tabular}
}
\end{table}

\begin{table}[h!]
\caption{Architecture's 4 {(MRMR)} model performance with multiple quantum Ridge output instances, executed on \textbf{noisy-model} simulators.}
\label{table:tabC12}
\addtolength{\tabcolsep}{-0.26em}
\resizebox{\columnwidth}{!}{%
\begin{tabular}{c|c|ccc|ccc|ccc}
\hline
 & & \multicolumn{3}{c|}{\textbf{Classical Reservoir}} 
 & \multicolumn{6}{c}{\textbf{Quantum Reservoir}} \\
\hline
 & & \multicolumn{3}{c|}{Simulation}
 & \multicolumn{3}{c|}{Simulation}
 & \multicolumn{3}{c}{Noisy} \\
\hline
\textbf{Res} & \textbf{Out-Num}
 & MAE & RMSE & $R^2$
 & MAE & RMSE & $R^2$
 & MAE & RMSE & $R^2$ \\
 \textbf{Neur} & \textbf{x} & & & & & & & & & \\
  & \textbf{Q-Out} & & & & & & & & & \\
\hline
2x5  & 2x5  & 0.0578 & 0.0724 & 0.8006 & 0.0206 & 0.0298 & 0.9695 & 0.0215 & 0.0307 & 0.9623 \\
2x10 & 2x10 & 0.0369 & 0.0497 & 0.9058 & 0.0195 & 0.0271 & 0.9712 & 0.0203 & 0.0294 & 0.9648 \\
2x20 & 2x10 & 0.0146 & 0.0240 & 0.9762 & 0.0168 & 0.0237 & 0.9805 & 0.0171 & 0.0236 & 0.9811 \\
3x5  & 3x5  & 0.0250 & 0.0338 & 0.9573 & 0.0198 & 0.0275 & 0.9701 & 0.0202 & 0.0291 & 0.9653 \\
3x10 & 3x10 & 0.0245 & 0.0349 & 0.9551 & 0.0176 & 0.0249 & 0.9791 & 0.0185 & 0.0262 & 0.9724 \\
3x20 & 3x10 & 0.0144 & 0.0205 & 0.9846 & \textbf{0.0141} & \textbf{0.0197} & \textbf{0.9852} & 0.0167 & 0.0215 & 0.9817 \\
5x5  & 5x5  & 0.0488 & 0.0593 & 0.8598 & 0.0185 & 0.0263 & 0.9782 & 0.0179 & 0.0246 & 0.9793 \\
5x10 & 5x10 & 0.0327 & 0.0461 & 0.9193 & 0.0163 & 0.0240 & 0.9810 & 0.0168 & 0.0228 & 0.9822 \\
\hline
\end{tabular}
}
\end{table}


\bibliographystyle{IEEEtran}
\bibliography{sn-bibliography}

\end{document}